\documentclass[final,5p,times,twocolumn]{elsarticle}

\usepackage{amssymb}
\usepackage{amsmath}

\usepackage{hyperref}
\usepackage[capitalise]{cleveref}
\crefname{appendix}{}{} \usepackage{booktabs}
\usepackage{siunitx}
\usepackage{multirow}
\usepackage{subcaption}
\usepackage{bm} \usepackage{dsfont} \usepackage{algpseudocode}
\usepackage[ruled,vlined]{algorithm2e}
\usepackage{snapshot}

\usepackage{tikz}
\usetikzlibrary{arrows.meta,bending,positioning,angles,quotes}
\usetikzlibrary{colorbrewer}
\usetikzlibrary{calc}
\usetikzlibrary{patterns}
\usetikzlibrary{decorations}
\usetikzlibrary{external}
\usetikzlibrary{positioning}
\usepackage{pgfplots}
\pgfplotsset{compat=1.16}
\usepgfplotslibrary{colorbrewer}
\usepgfplotslibrary{fillbetween}
\usepgfplotslibrary{groupplots}
\usepgfplotslibrary{statistics}
\usepgfplotslibrary{external}
\usepackage[l3]{csvsimple}

\definecolor{darkgreen}{rgb}{0,,0.5,0}

\newcommand{\overbar}[1]{\mkern1.5mu\overline{\mkern-1.5mu#1\mkern-1.5mu}\mkern 1.5mu}

\newcommand{\flexi}{FLEXI}

\newcommand{\relexi}{Relexi}

\journal{}

\begin{document}

\begin{frontmatter}

\title{A Provably Robust Multi-Jet Framework applied to Active Flow Control of an Airfoil in Weakly Compressible Flow.}

\author[label1]{Rohan Kaushik\corref{cor1}}
\author[label1]{Anna Schwarz}
\author[label1]{Andrea Beck}

\affiliation[label1]{organization={Institute of Aerodynamics and Gas Dynamics, University of Stuttgart},
            addressline={Wankelstraße 3},
            postcode={70563},
            city={Stuttgart},
country={Germany}}

\cortext[cor1]{Corresponding author}

\begin{abstract}
Reinforcement learning has by now become well established in finding excellent flow control strategies for a variety of scenarios.
Existing literature has focused on using a simple two-jet solution (and variants there-of) or a straightforward mean-centered multi-jet setup.
This mean-centering approach is however non-injective in nature, such that distinct action predictions by the actor network can lead to the same implemented jet-intensities.
Thus, the potential of true multi-jet setups still remains unexplored.
To this end, in this study we first theoretically analyze multi-jet setups, highlighting the aforementioned pitfall and offer a viable alternative.
We also derive upper-bounds on the \textit{running costs} of these setups, and find the proposed approach to have a jet-count-independent maximum running cost (compared to a near-linear scaling for the traditional setup).
The mean-centered and proposed multi-jet setups are applied to a variety of flow-configurations, to test performance and learning capabilities. We use the flow solver \flexi{}, based on the discontinuous Galerkin method, for solving the compressible Navier-Stokes-Fourier set of equations and the \relexi{} package for the RL related aspects of this work.
The new formulation proves effective in learning more complex flow-control strategies, coordinating the jets in a sophisticated manner so as to produce favorable outcomes at minimal actuation cost.
For the cylinder-in-channel case, this results in drag and total-force suppression to beyond an idealized symmetric case, whereas for the airfoil the separation region is minimized and significant improvements in aerodynamic efficiency are observed (from $53\%$ up to $73\%$ depending on jet configuration). Additionally, we also incorporate some best practices from traditional RL literature to show fast, reproducible and reliable learning, thereby bringing down the upfront training costs.
This study thus provides a robust and mathematically grounded approach to multi-jet design and closes a hitherto overlooked theoretical gap.
\end{abstract}


\end{frontmatter}

\section{Introduction}

Controlling the flow over an immersed body is a fundamental industrial activity with numerous practical benefits.
Applications of such systems range from improving aerodynamic characteristics in transport vehicles to improving mixing efficiencies in chemical reactors \cite{Gad-el-Hak_2000}.
Traditional methods have focused largely on \textit{passive} flow control systems involving specialized geometry modifications, essentially `baking in' the flow-control device into the machines' design \cite{mclellan1988history,beratlis_separation_2017,bechert_viscous_1989,Chambers2003ConceptTR,lin2002review}.

\textit{Active} flow control (AFC) on the other hand involves expending energy to effect changes within the flow in consideration.
These can take many forms, from methods such as direct momentum injection via jet actuators, to periodic forcing techniques and plasma actuators for flow acceleration \cite{amitay1998aerodynamic,greenblatt_control_2000,kametani_direct_2011,yousefi_three-dimensional_2015,voevodin_improvement_2019,desalvo2012high,wu1998post,radespiel2016active}.
Having started from open-loop control, i.e. AFC systems with no feedback from the controlled system, the field gradually moved towards closed-loop control, with controllers adjusting their behavior based on the modified flow.
\citet{warui1996feedback} were likely the first to apply a direct proportional control law to suppress vortex shedding on a cylinder experimentally, while \citet{min1999suboptimal} soon after utilized an invasive adjoint based optimization approach to achieve the same, both bearing excellent results.
\citet{muddada_active_2010} used a similar closed-loop approach, instead relying on a variant of integral control to achieve impressive drag reduction on a low Reynolds number cylinder-in-channel flow.
Later studies extended this approach to other test cases (such as multi-foils and turbine cascades) as well as more advanced techniques such as PID, model-predictive and sliding-mode control laws \cite{lee2013closed,michel2024novel,kaul2022active,doi:10.2514/1.J055728,staats2017closed, krentel2010application}.

The main challenge with these advanced controllers is the high degree of non-linearity and chaos inherent in the Navier-Stokes equations, requiring high dimensional control spaces and/or simplifications via specialized knowledge.
Deep Reinforcement Learning (DRL) being a model-free approach with a track record of finding intricate solutions, thus seemed poised to take up the challenge.
The effectiveness of DRL in controlling and suppressing drag was first explored by \citet{rabault2019artificial} in their seminal contribution applying it to reducing drag on a 2D cylinder placed in a channel using a pair of synthetic jets.
The jets were forced to have a net zero injected mass flow rate to curb excess momentum injection, implemented by predicting only a single jet's intensity and setting the other to be its negative value.
This was followed by a series of studies investigating the effectiveness of DRL in controlling various aerodynamic quantities across setups ranging from airfoils to channels, all with remarkable success \cite{rabault2019accelerating,rabault2020deep,tang2020robust,wang2022deep,GARCIA2025109913,suarez2023active,suarez2024flow,suarez2025active}.

The blowing/suction approach via zero net mass flow rate synthetic jets is by far the most popular approach in controlling flow on immersed bodies.
\citet{wang2022deep} used it for AFC on airfoils in low Reynolds number flows, \citet{GARCIA2025109913} for controlling separation and \citet{mondal2025shocks} for mitigating shocks.
The initial studies focused on two-jet systems as described earlier, where enforcing a zero net mass flow rate condition is trivial.
A few studies explored the effects of using more than two jet actuators, such as using multiple pairs and/or compensator jets \cite{mondal2025shocks,suarez2024flow,suarez2025flow}, or using 3-4 jets in a mean-centering setup \cite{tang2020robust,wang2022deep,GARCIA2025109913}.
The first approach is a straightforward extension of the two-jet system, and results in `locking in' the jet response to the jet positions because of the user's bias in picking which jet positions form a pair (or the compensator jet).
The latter allows for an arbitrary number of jets, with the DRL agent controlling each individually by producing as many outputs (one for each jet).
However, this approach has an even deeper problem, in that it leads to a non-injective mapping from the DRL agent's outputs to the implemented jet-intensities.
A mean-centering operation by definition implies that two distinct input vectors, if spread similarly about their respective mean-values, will result in the same outputs upon its application.

This has significant implications for the real-world deployment of such systems, as the mean-centering modulation could collapse two different agent predictions onto a single jet strength distribution, leading to undesirable and ambiguous results.
Additionally, we believe this might limit the agent's ability to explore and learn effectively (as discussed in \cref{sec:znmf_tradjet_theory}), since in previously cited works employing mean-centering the agents invariably settle upon simplistic near-constant blowing/suction strategies.

Another issue when dealing with DRL in fluid dynamics is a lack of repeatability studies.
A core tenet of all machine learning research is the recognition that these methods are stochastic in nature and thus, to make generalizable claims one should test different random initializations to rule out chance effects.
This ensures that observed performance is robust rather than an artifact of a particular seed or lucky configuration.
This void is in part due to the fact that compute costs of each individual simulation can be quite high, thereby raising the training costs many-fold.
To this end, we also discuss and implement certain standard RL practices such as learning rate warm-up, and demonstrate fast, reliable and repeatable training.
These measures are standard in DRL research, but have so far not been applied in the DRL-AFC context.

Thus, in this study we propose an alternative injective approach for dealing with multiple ($N>2$) jets, from which the traditional two-jet case emerges as a special case for $N=2$.
Training is observed to be highly consistent across different initializations, and maximum \textit{costs} (as defined up ahead) are found to be more economical than the mean-centering case.

The paper is organized as follows.
\cref{sec:methodology} details the simulation setups used in this study, along with the reinforcement learning algorithm and the particular modifications/improvements used therein.
\cref{sec:methodology-multijets} describes the different approaches to implementing zero net mass flow rate multi-jets, first outlining the traditional method and its pitfalls, before presenting our alternative.
\cref{sec:results} discusses the results of the applied strategies to the cylinder-in-channel and airfoil-in-channel setups, across jet numbers and configurations.
Finally, \cref{sec:conclusion} wraps up and summarizes the contributions of this study, while \cref{sec:validation,sec:app-cost,sec:app-airfoiljetpos} contain additional analyses.
\section{Methodology}
\label{sec:methodology}

We first describe the simulation environment for the two test setups, and then the reinforcement learning setup along with some implementation details. All reported quantities are suitably non-dimensionalized.

\subsection{Simulation Setup}
\label{sec:methodology-simsetup}

\subsubsection{Cylinder-in-Channel}
\label{sec:methodology-simsetup-cylinder}

\begin{figure*}[t]
    \centering
    \tikzsetnextfilename{fig_simulation_setup_cylinder}
    \includegraphics{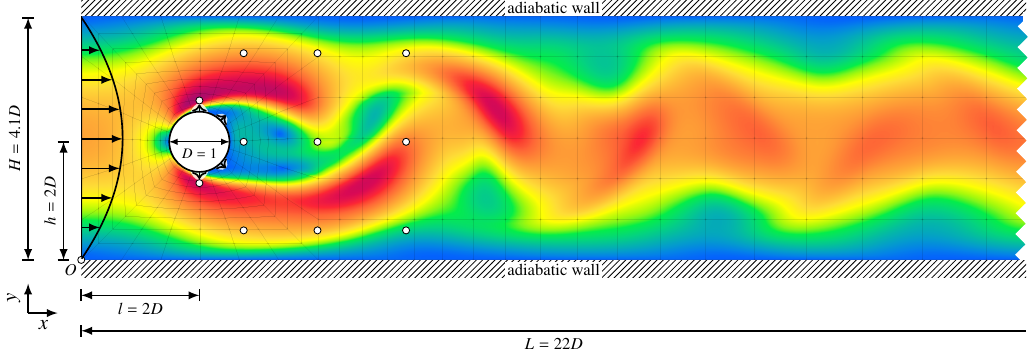}
    \caption{Simulation setup for the flow around a two-dimensional cylinder in a channel. The positions of the 11 pressure probes are highlighted by white circles and the jets by white areas around the cylinder. The field solution shows the velocity magnitude. The domain is clipped here for better visual representation.}
    \label{fig:simulation_setup_cylinder}
\end{figure*}

\begin{figure}[t]
    \centering
    \tikzsetnextfilename{fig_jet_setup_cylinder}
    \includegraphics{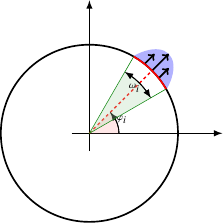}
    \caption{Jet setup on the cylinder, with the geometric parameters for the $i^{th}$ jet having been plotted. Here ${\phi}_i$ is the angular position of the jet-center, and $\omega_i$ the angular-width of the jet-slot. The slot opening is marked in red and the dimensions exaggerated for better visualization.}
    \label{fig:jet_setup_cylinder}
\end{figure}

The cylinder-in-channel environment is well tested by now and the simulation setup from \citet{KURZ2025106854} is utilized here, which itself is in line with those described in \cite{rabault2019accelerating,rabault2019artificial}. At its core, it consists of a 2D channel with a cylinder placed at its centerline. The upper and bottom walls of the channel are adiabatic in nature, with no-slip velocity boundary conditions. The cylinder wall is the same adiabatic, no-slip wall, except for the sections defining the jets. The setup is shown in \cref{fig:simulation_setup_cylinder}. A parabolic velocity profile is imposed at the inlet of the setup
\begin{align}
    U(y) = U_m \frac{4y(H-y)}{H^2}, \label{eq:meth-simsetup-cyl-inflowvel}
\end{align}
with $H=4.1D$ being the channel height, $U_m$ the maximum velocity at the centerline and $D$ the diameter of the cylinder.
The Reynolds number ($Re$) and Mach number ($Ma$) are defined with respect to the mean inlet velocity ($\overbar{U}$)
\begin{align}
    Re = \frac{\overbar{U} D}{\nu}, \ \
    Ma = \frac{\overbar{U}}{c_{\infty}}, \ \
    \overbar{U} = \frac{2}{3} U_m, \label{eq:meth-simsetup-cyl-ReMaUbar}
\end{align}
with $\nu$ being the kinematic viscosity and $c_{\infty}$ the freestream speed of sound.
At the inlet, the density and pressure are set as constants $\rho(y) = \rho_{\infty}$ and $p(y) = p_{\infty} = \frac{\rho_{\infty}}{\kappa}\left( \frac{\overbar{U}}{Ma} \right)^2$, $\kappa$ being the heat capacity ratio.
At the outlet, a constant pressure-outflow condition from \citet{carlson2011FUN3D} is imposed to recover $p_{\infty}$.
In this work, $Re=100$ and $Ma=0.2$ are used to conduct the cylinder simulations.
One can notice that the cylinder is offset from the centerline by $5\%$ of the diameter $D$ to hasten the onset of the vortex shedding regime, similar to \citet{rabault2019artificial}.

The synthetic jets' positions are specified by two quantities, the angular position of the jet centerline ($\varphi_i$) and the angular jet-width ($\omega_i$).
The jets are implemented such that they introduce a mass flux component ($j_{\perp}$) perpendicular to the surface within their specified regions.
This normal mass flux component is modeled as a cosine distribution, reaching its peak at the center of the jet-slot and zero at its ends
\begin{equation}
    j_{\perp}(x, y) = \begin{cases}
        Q_i \frac{\pi}{\omega_i D} \cos\left(\frac{\pi}{\omega_i} (\varphi - \varphi_i) \right) & \text{if} \ |\varphi - \varphi_i| \leq \frac{\omega_i}{2}, \\
        0 & \text{\small{otherwise}},
    \end{cases} \label{eq:meth_cyl_jetdist}
\end{equation}
where $\varphi(x,y)$ is the angle of the point $(x,y)$ with respect to the cylinder's center and the $x$-datum (see \cref{fig:jet_setup_cylinder}).
Integrating this across the jet-slot, $Q_i$ becomes the total mass flow rate being ingested/ejected by the $i^{th}$ jet.
For further details on the setup readers are referred to the earlier cited studies.

\subsubsection{Airfoil-in-Channel}
\label{sec:methodology-simsetup-airfoil}

\begin{figure*}[t]
    \centering
    \tikzsetnextfilename{fig_simulation_setup_airfoil}
    \includegraphics{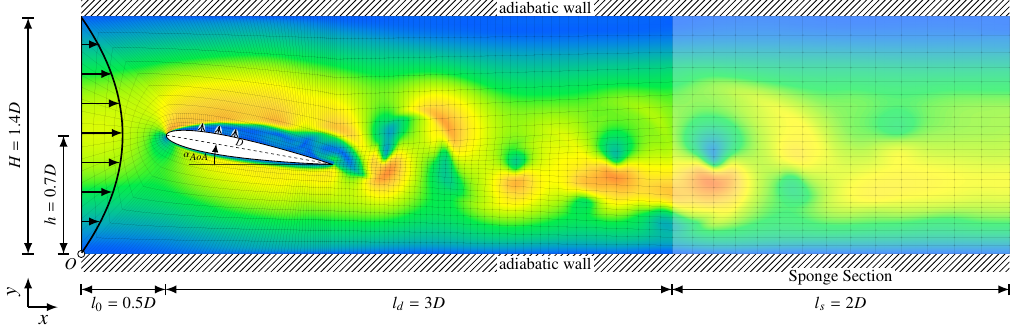}
    \caption{Simulation setup for the flow around a two-dimensional airfoil in a channel. The field solution shows the velocity magnitude for $Ma=0.4$, and the jet positions are visible on the suction surface. The highlighted section at the back represents the numerical sponge, and its effectiveness can be gleaned from the almost completely damped wake-vortices at the exit-boundary.}
    \label{fig:simulation_setup_airfoil}
\end{figure*}

\begin{figure}[t]
    \centering
    \tikzsetnextfilename{fig_jet_setup_airfoil}
    \includegraphics{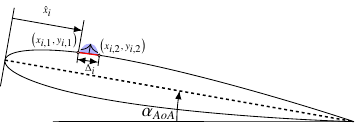}
    \caption{Jet setup on the airfoil, with the geometric parameters for the $i^{th}$ jet having been plotted. Here $\hat{x}_i$ is the chord-wise beginning position of the jet, and $\Delta_i$ the width of the jet-slot. $\left(x_{i,1},y_{i,1}\right)$ and $\left(x_{i,2},y_{i,2}\right)$ are the starting and ending positions of the jet, computed from the specifications of $\hat{x}_i$ and $\Delta_i$. The slot opening is marked in red and the dimensions exaggerated for better visualization.}
    \label{fig:jet_setup_airfoil}
\end{figure}

The simulation environment used here is the same as that from \citet{wang2022deep} and \citet{GARCIA2025109913}, wherein a NACA0012 airfoil is situated in a channel at an angle of attack $\alpha_{AoA} = 10^{\circ}$ with synthetic jets placed along its surfaces.
The setup is shown in \cref{fig:simulation_setup_airfoil}, where the channel height is $H = 1.4 D$ and the channel length is $3.5 D$ ($D$ being the chord-length of the airfoil).
The upper and lower walls of the channel as well as the airfoil surface (barring the jet sections) are treated as adiabatic no-slip walls.
As before, the boundary on the left is treated as the inflow with a parabolic velocity profile (\cref{eq:meth-simsetup-cyl-inflowvel}), and a constant pressure condition is imposed at the outflow on the right to recover $p_{\infty}$.
The airfoil's leading edge is placed on the channel centerline at a distance of $l_0=0.5D$ from the channel inlet, i.e. $(x, y)_{LE} = (l_0, H/2)$.
To aid in numerical stability at the outflow, a numerical \textit{sponge} section is added after the channel, starting from $x=3.5D$ and stretching till $x=5.5D$ \cite{krais2021flexi}.
It forces the flow-state to a quasi-equilibrium using an additional source term based on the moving average, and its purpose here is to dampen the exiting wake vortices so as to prevent instabilities at the constant-pressure outlet of the domain.
The Reynolds number and Mach number are defined with respect to the mean inflow velocity (\cref{eq:meth-simsetup-cyl-ReMaUbar}), and set to $Re=3000$ and $Ma=0.4$. The flow in these conditions is separated, as is clearly observable in \cref{fig:simulation_setup_airfoil} and the $C_P$ plots in \cref{fig:af_cp_lineplot-3and6jets}.

The synthetic jets' positions are specified by three quantities, their beginning position along the chord-line of the airfoil ($\hat{x}_i$), the jet-width ($\Delta_i$) and the surface (suction/pressure) on which they are to be located. We assume the jet-width is small enough to neglect surface curvature. The jet-slot extends from $(x_{i,1}, y_{i,1})$ to $(x_{i,2}, y_{i,2})$ on the airfoil surface, and these quantities are computed a-priori to a simulation based on $\hat{x}_i$ and $\Delta_i$ (see \cref{fig:jet_setup_airfoil}). Then, the normal mass flux component ($j_{\perp}$) introduced by the jet is modeled by a sinusoidal distribution, reaching its peak at the center of the jet-slot and zero at its ends
\begin{equation}
    j_{\perp}(x, y) = \begin{cases}
        Q_i \frac{\pi}{2 \Delta_i} \sin\left(\frac{\pi (x-x_{i, 1})}{\Delta_i \cos (\phi_i)}\right) & \text{\small{if} } \ x_{i,1} \leq x \leq x_{i,2} \ \ \& \\
            & (x,y) \in \text{\small{correct surface}}, \\
        0                                                                                  & \text{\small{otherwise}.}
    \end{cases} \label{eq:meth_af_jetdist}
\end{equation}
The term `correct surface' refers to the specified suction/pressure surface that the $i^{th}$ jet lies on, and
\begin{align}
    \phi_i = \arctan \left( \frac{y_{i,2} - y_{i,1}}{x_{i,2} - x_{i,1}} \right)
\end{align}
is the angle of the jet-slot w.r.t. the $x$-datum of the mesh. As before, integrating along the jet, $Q_i$ becomes the total mass flow rate of the $i^{th}$ jet.

\subsubsection{Flow Solver and Mesh Details}
\label{sec:methodology-simsetup-solver}

The simulations are carried out using \flexi{}  \cite{krais2021flexi}, a scalable high-order accurate flow solver for hyperbolic and parabolic partial differential equations, with a focus on compressible flows.
It employs the discontinuous Galerkin spectral element method (DGSEM), works across structured and unstructured hexahedral meshes and has been extensively validated \cite{blind2022performance,blind2024wall,durrwachter2021efficient}.

For the cylinder-in-channel setup, the domain is discretized into $372$ quadrilateral elements using a block-structured, fourth-order curved mesh representation.
For the airfoil setup, it is discretized into $3952$ quadrilateral elements using second-order polynomials for defining element geometries.
The simulations are carried out on a collocated Legendre-Gauss interpolation and integration strategy.
Based on the validation study in \cref{sec:validation}, a polynomial order of $N=4$ is used for the cylinder setup, resulting in $9300$ degrees of freedom.
Similarly, a polynomial order of $N=3$ is used for the airfoil setup, resulting in $63232$ degrees of freedom.
Since the flow is compressible, a Mach number has to be imposed to recover the background pressure, and hence the earlier imposition of $Ma=0.2$ and $Ma=0.4$ for the cylinder and airfoil respectively.

Evaluation metrics include the lift, drag, total force and pressure coefficients
\begin{align}
    C_L = \frac{F_y}{\overbar{U}^2 D \rho_{\infty} / 2},
    C_D = \frac{F_x}{\overbar{U}^2 D \rho_{\infty} / 2}, \nonumber \\
    C_F = \frac{\sqrt{F_x^2+F_y^2}}{\overbar{U}^2 D \rho_{\infty} / 2}, C_P = \frac{p-p_{\infty}}{\overbar{U}^2 D \rho_{\infty} / 2}
\end{align}
where $F_y$, $F_x$ are the forces on the surface of the immersed body in the $y$ and $x$ directions respectively. All results are reported in the non-dimensionalized time $t^* = t / t_{ref}$ with $t_{ref} = D / \overbar{U}$.

\subsection{Reinforcement Learning}
\label{sec:methodology-rl}

\begin{figure}[t]
    \centering
    \tikzsetnextfilename{fig_rl_env}
    \includegraphics{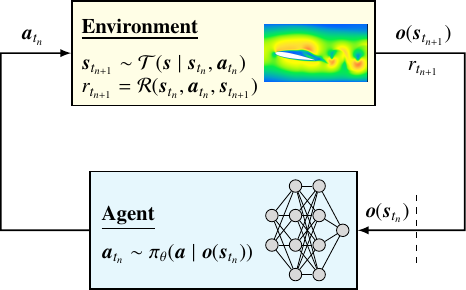}
    \caption{Agent-environment interaction in a DRL setup. $\mathcal{T}(\boldsymbol{s} \mid \boldsymbol{s}_{t_n}, \boldsymbol{a}_{t_n})$ are the theoretical transition probabilities dictating state evolution and $\pi_\theta(\boldsymbol{a} \mid \boldsymbol{o}(\boldsymbol{s}_{t_n}))$ the probability distributions defining the actions (remaining symbols as defined in \cref{sec:methodology-rl}).}
    \label{fig:fig_rl_env}
\end{figure}

Reinforcement learning is a set of supervision-free methods for learning optimal behavioral policies for Markov Decision Processes (MDPs).
MDPs are a class of sequential interaction tasks, where each interaction involves a state observation, motivating an action which advances the system into a new state, leading to a reward based on its optimality ($\boldsymbol{o}(\boldsymbol{s}_{t_{n}}), \boldsymbol{a}_{t_{n}}, \boldsymbol{s}_{t_{n+1}}, r_{t_{n+1}}$, respectively, see \cref{fig:fig_rl_env}).
The cycle is then continued-on from the new state, either until a terminal point or until infinity (in theory).
The goal is to find decision policies that maximize the cumulative rewards of such interactions.
Deep reinforcement learning (DRL) is a sub-branch which utilizes neural-networks to predict these actions.
Within DRL there exist a plethora of approaches \cite{schulman2017proximal,lillicrap2015continuous,barth2018distributed,gu2016continuous}, and of them the Proximal Policy Optimization (PPO) algorithm is used for this study.

PPO has two main advantages - first that it yields competitive performance at easily manageable computational and mathematical complexity, and second that it is widely in use with well-established tools and literature to support.
PPO is an episodic method, meaning that learning occurs only after the environment reaches a terminal state.
At that point, all data collected during the episode is processed and used for training.
Consequently, the algorithm optimizes the policy only within the duration of each episode. For this reason, the episode length must be sufficiently long to enable the agent to learn policies that remain stable and effective over extended time horizons.

PPO involves the simultaneous training of two neural networks, a \textit{policy} (or \textit{actor}) network and a \textit{value} (or \textit{critic}) network.
Both networks use the observations $\boldsymbol{o}_{t_{n}} = \boldsymbol{o}(\boldsymbol{s}_{t_{n}})$ as inputs.
The actor network learns the control policy and outputs the actions $\boldsymbol{a}_{t_{n}}$ that are applied to the environment.
In parallel, the critic network estimates the \textit{discounted cumulative future rewards} associated with a given state.
This estimate is incorporated into the policy loss calculation to reduce variance and improve the stability of the learning process.
Further details about PPO can be found in \cite{schulman2017proximal,sutton2020reinforcement}.

\subsubsection*{Observations $\boldsymbol{o}_{t_n}$}

While the full state $\boldsymbol{s}_{t_n}$ corresponds to the complete flow-field at time instant $t_n$, the networks only operate on partial \textit{observations} $\boldsymbol{o}_{t_n}=\boldsymbol{o}(\boldsymbol{s}_{t_n})$ of this whole picture.
The trade-off lies in constructing these partial observations such that they provide enough meaningful information about the state to the actor/critic networks while keeping the input size small and manageable.
In the cylinder setup, this amounts to pressure difference values $\Delta p_i = p(\boldsymbol{x}_{\text{probe,i}})-p_{\infty}$ measured at the 11 highlighted probes in \cref{fig:simulation_setup_cylinder}.
This is in-line with previous studies such as \citet{KURZ2025106854}.
The position of these probes is kept the same as those found in \cite{KURZ2025106854,rabault2019accelerating}, which were chosen based on the researchers' personal experiences.

To remove this personal bias, in the airfoil setup probe positions are based on a heuristic procedure outlined in \cref{sec:app-airfoiljetpos}.
The crux of it is that one observes quantities-of-interest across numerous probe locations and a large time scale, and rejects probe locations with high correlations.
This leaves only positions with distinct information content, thereby striking a balance between the number of probes and the information they contain.

\begin{figure}[t]
    \centering
    \tikzsetnextfilename{fig_RPpos_airfoil}
    \includegraphics{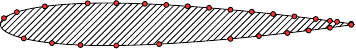}
    \caption{Probe positions on the airfoil, used to record observations and pass along as input to the policy and value networks. Positions are determined based on the method described in \cref{sec:app-airfoiljetpos}.}
    \label{fig:RPpos_airfoil}
\end{figure}

In practical applications, it would be unlikely for a deployed AFC system to have access to data from precise points in the wake (consider for example, a system deployed on an aircraft's wing).
It is far more likely to have access to probes distributed on the body's surfaces. To this end, only probe positions on the surface of the airfoil are considered. This results in 28 positions distributed across the suction and pressure surfaces, as shown in \cref{fig:RPpos_airfoil}.
This definitely limits the exposure of the policy/value networks to the full state of the flow field, having to operate with no access to wake states.
This should be manageable in the present sub-sonic flow scenario, since atleast one of the characteristics should be moving against the general flow velocity thereby allowing for information from the wake to travel upstream \cite{lax1973hyperbolic,toro2013riemann}.
To our knowledge this is the first study to explore DRL-AFC in such a restricted information application, and a good test case of the constraints in which a real-world system might have to function.
Additionally, here the pressure difference values are normalized using the dynamic pressure to turn them into coefficients of pressure ($C_P$) values, before passing them on to the networks.
The time-derivative of these $C_P$ values at the probe positions is also supplied alongside, to enhance the agent's awareness of the flow-field using temporal data.

\subsubsection*{Actions $\boldsymbol{a}_{t_n}$}
\label{sec:meth-rl-actions}

Learning a policy in PPO relies somewhat on a stochastic process.
To ensure a variety of experiences to learn from, the actor network is tasked with predicting a probability distribution of actions.
During the training phase, the actions at each time step are drawn according to these probability distributions.
Commonly used distributions are those that are easily parameterizable (for example the Gaussian distribution, depending only on the mean $\mu$ and variance $\sigma^2$).
Then the task of the actor network boils down to predicting these defining parameters, for every given observation.
However, to ensure consistency during evaluation, actions aren't drawn probabilistically.
Rather, the maximum likelihood value (or the mode) of the predicted distribution is used as the action value.
This strategy gives an easy melding of both worlds, stochasticity to ensure variety in training and determinism to ensure consistency in evaluation.

Based on the framework detailed in \cref{sec:methodology-simsetup}, for controlling $N$ jets, the agent needs to supply $N$ individual $Q_i$ values (it being the mass flow rate through the $i^{th}$ jet, as defined in \cref{,eq:meth_cyl_jetdist,eq:meth_af_jetdist}).
To ensure stability and eliminate the need for a dedicated reservoir to manage excess/deficit momentum, two additional conditions are imposed: \begin{itemize}
    \item $|Q_i| \leq Q^{max}$ (capped individual total mass flow rates),
    \item $\sum_{i=1}^N Q_i = 0$ (zero net mass flow rate).
\end{itemize}

In this study, the outputs of the actor network $\boldsymbol{a}_{t_n}$ are bounded to the range $[0, 1]$, and passed on to a modulating function $\boldsymbol{f}(\boldsymbol{a}_{t_n})$ which scales them to the range $[-1, 1]$ while enforcing $\sum_i^N f_i(\boldsymbol{a}_{t_n}) = 0$.
The implemented jet-intensities are then computed as $Q_i = Q^{max} f_i(\boldsymbol{a}_{t_n})$, thereby fulfilling these two conditions.
The exact imposition of the zero net mass flow rate condition for $N>2$ is not trivial, and the traditional and proposed approaches are discussed in detail in \cref{sec:methodology-multijets}.

To prevent simulation instabilities, it is essential to ensure a smooth temporal transition from one action to the next \cite{rabault2019artificial}.
In support of this goal, implemented jet intensities are smoothly blended as:
\begin{align}
    Q_i(t) = Q_i(t_n) + \left(Q_i(t_{n+1}) - Q_i(t_n)\right)\left(1 - e^{-\zeta(t-t_n)} \right), \\
    t \in [t_n, t_{n+1}]. \nonumber
\end{align}
The parameter $\zeta \left[\frac{1}{s}\right]$ controls the rate of change of the additive term, and must be chosen such that it guarantees transition to $Q_i(t_{n+1})$ by the time $t=t_{n+1}$ but in a smoothly controlled manner.

\subsubsection*{Reward $r_t$}

The reward $r_{t_{n+1}} = \mathcal{R}(\boldsymbol{s}_{t_{n}}, \boldsymbol{a}_{t_{n}}, \boldsymbol{s}_{t_{n+1}})$ is a scalar value computed using the reward function $\mathcal{R}(\cdot)$.
The reward signals to the agent whether its suggested action(s) leads to favorable states and time-lines or not.
Of interest is not just the immediate consequence of an action, but also the effect it has on the future.
To take this into account, \textit{discounted cumulative future rewards} (DCFR) are computed and used for training. They are defined as
\begin{align}
    \text{DCFR}_{t_{n+1}} = \sum_{i=0}^{\infty} \gamma^i r_{t_{n+1+i}},
\end{align}
with $\gamma<1$ being a weight used to assign importance to future rewards, often called the \textit{discount factor}.
Broadly speaking, the aim of all DRL methods is to maximize this quantity\footnotemark, see \cite{schulman2017proximal,sutton2020reinforcement,lillicrap2015continuous}.
\footnotetext{Strictly speaking, the expected `return' of a trajectory is maximized, which is the \textit{undiscounted} sum of rewards. However, in practice DRL algorithms often use the DCFR in defining the value functions and networks.}

Two different reward functions are used for the two simulation setups, given that both have differing objectives. For the cylinder case, the objective is to reduce drag and the total forces on the cylinder, and so the reward function used is from \citet{KURZ2025106854}, given as
\begin{align}
    r_{t_{n}} = \frac{\langle C_D \rangle^{unactuated} - C_{D}(t_n)}{\langle C_D \rangle^{unactuated} - C_{D,min}} - \chi_1 |C_L(t_n)|,
    \label{eq:meth-reward-cyl}
\end{align}
with $\langle \cdot \rangle$ denoting the average of a quantity, $C_{D,min}$ the estimated minimum achievable drag and $\chi_1$ a parameter determining the strength of the $C_L$ penalty.

For the airfoil case, the goal is to increase the aerodynamic efficiency $C_L/C_D$, and the so reward function from \citet{GARCIA2025109913},
\begin{align}
    r_{t_n} =  \chi_1 \left( \frac{C_L(t_n)}{C_D(t_n)} + \chi_2 \right)
    \label{eq:meth-reward-airfoil}
\end{align}
is used. Here $\chi_1$ and $\chi_2$ are constants to help scale the rewards to more manageable ranges. In this study, they have been set to
\begin{align*}
\chi_1 &= \frac{1}{3 \ \text{stddev}\left[ C_L/C_D \right]^{unactuated} }, \\
    \chi_2 &= - \left( \langle C_L/C_D \rangle^{unactuated} + \text{stddev}\left[ C_L/C_D \right]^{unactuated} \right).
\end{align*}
where $\text{stddev}\left[\cdot\right]$ represents the standard deviation of a quantity.

\subsubsection{Policy and Value Networks}

The neural networks used in this work are simple multi-layer perceptrons, or MLPs.
These are just the concatenation of a series of matrix multiplications with (non-linear) activation functions sandwiched in between.
Both the policy and value networks consist of two hidden layers 256 neurons wide, using hyperbolic tangent as the activation function in the cylinder's case and the Gaussian error linear unit (GeLU) function in the airfoil's case.
In the value networks, a final layer is used to take the 256 dimensional representation down to a single scalar value, along with a hyperbolic tangent activation and a final weight to adjust as needed.
In the policy networks, the output of the hidden layers is piped through two separate paths, both taking the hidden representation down to $N$ or $N-1$ values (depending on the multi-jet framework, see \cref{sec:methodology-multijets}).
One path represents the auxiliary variable ($\nu$ or $\sigma$, see \cref{sec:meth-multijet-rlnotes}), and these values are further capped by a sigmoid activation function, finally being scaled/translated to the range $[\text{auxvar}_{min}, \text{auxvar}_{max}]$.
The other path represents the modes of the probability distribution.
In the traditional mean-centered approach a final sigmoid activation is applied to cap the outputs to $\left[0, 1\right]$, while in the newly proposed approach a softplus activation ensures positive outputs, before tangling them up according to \cref{eq:meth-newapp-fi}.

\subsubsection{Implementation Details}

The python package \relexi{} \cite{kurz2022relexi} is used to handle the machine learning aspects of this work, as well as for management and creation of the various simulation environments. \relexi{} is designed to couple scalable HPC simulations to a reinforcement learning framework based on the tensorflow library TF-Agents \cite{tensorflow2015-whitepaper,TFAgents} using smartsim \cite{partee2022using}. It automates workload distribution and the handling of simultaneous instances of parallelized environments across available compute resources.

\begin{figure}[t]
    \centering
    \tikzsetnextfilename{fig_learningrate}
    \includegraphics{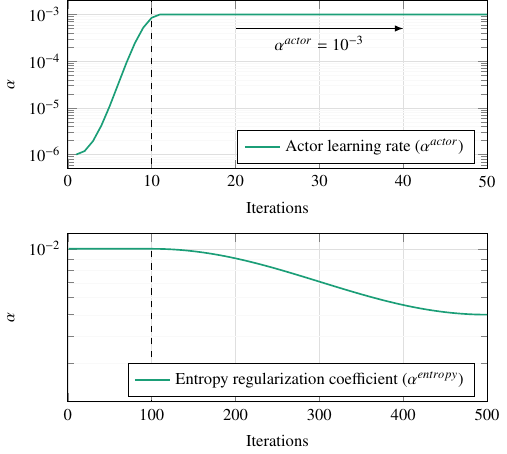}
    \caption{Learning rate and entropy penalty coefficient evolution throughout training.}
    \label{fig:fig_learningrate}
\end{figure}

Like \citet{KURZ2025106854}, the RL environment is provided with a set of different restart files to start the simulations from, ensuring a variety of state-action-reward tuples for training.
The used PPO implementation also employs a Kullback-Leibler divergence based early stopping constraint \cite{KURZ2025106854,engstrom2019implementation} for the policy networks, where the KL divergence \textit{from} distribution $P$ \textit{to} distribution $Q$ is defined to be
\begin{align}
    D_{\mathrm{KL}}(P \parallel Q) = \int_{-\infty}^{\infty} p(x) \log \frac{p(x)}{q(x)} \, \mathrm{d}x.
\end{align}
It is a measure of \textit{distance} between the policy distributions from the beginning of an update iteration to the end of an update epoch, and policy network training is stopped if this value crosses a certain threshold $\mathcal{T}_{ES}^{actor}$.
We use the ADAM optimizer \cite{kingma2014adam} for carrying out the actual updates.
An entropy penalty term is also added to the loss function, to encourage exploration in the initial stages \cite{schulman2017proximal}.
Its coefficient is steadily decreased over time so as to enable the agent to focus more on refining the found polices (\cref{fig:fig_learningrate}).

As noted in numerous studies \cite{liu2025theoretical,abuduweili2024revisiting,kalra2024warmup,vaswani2017attention,goyal2017accurate,gotmare2018closer}, warming up the learning rate from a tiny value to the target can have a huge impact on model performance and training convergence.
This effect is magnified in optimizers with adaptive estimates like ADAM, primarily because of the lack of information in the initial update steps for the optimizer to build accurate estimates.
This is an issue faced by the authors firsthand, as training was noted to be particularly dependent on network weight initialization.
However, with the learning rate warmup schedule shown in \cref{fig:fig_learningrate}, training behaviour has been found to be quite consistent across different random initializations.

\begin{figure}[t]
    \centering
    \tikzsetnextfilename{fig_rl_staterecycling}
    \includegraphics{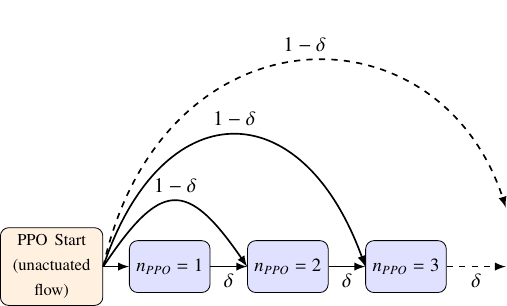}
    \caption{State recycling for PPO. Here, $n_{PPO}$ is the update iteration in the PPO algorithm, and $\delta$ is the fraction of statefiles that get passed on from iteration to iteration. At the beginning of the algorithm, all the statefiles are from the converged unactuated simulations.}
    \label{fig:fig_rl_staterecycling}
\end{figure}

Lastly, we also use the end-of-simulation state recyling strategy from \citet{suarez2025active}, labeled there as `DRL-10-s2'. The primary idea is that PPO being an episodic method, only learns policies producing optimal trajectories for simulations limited to the training episode duration. To ensure performant policy behaviour beyond these durations, at every PPO iteration the initial state-collection is altered to consist of a fraction $\delta$ corresponding to the final time-step of the previous iteration's simulations. The remaining $1-\delta$ fraction comprises the converged baseline cases that would otherwise have been used (see \cref{fig:fig_rl_staterecycling}).

\begin{table}[htb]
    \centering
    \begin{tabular}{lllp{3.25cm}}
        \toprule
        \textbf{Hyper-}         & \textbf{Cylinder}     & \textbf{Airfoil}     & \textbf{Description} \\
        \textbf{parameter}      & & & \\
        \midrule
        $t_{end}/t_{ref}$                       & 20                    & 8                    & Simulation end time. \\
        $\Delta t_{RL}/t_{ref}$                 & 0.25                  & 0.05                 & Action time interval. \\
        $\zeta$                 & \multicolumn{2}{c}{$5/\Delta t_{RL}$} & Action smoothing. \\
        $C_{D,\text{min}}$              & 2.6                   & -                    & \multirow{6}{3.25cm}{Reward defining parameters (see \cref{eq:meth-reward-cyl,eq:meth-reward-airfoil}).}\\
        $\langle C_{D} \rangle^{\dagger}$ & 2.9       & -                    &  \\
        $\langle C_L / C_D \rangle^{\dagger}$  & - & 2.895& \\
        $\text{std}\left[C_L/C_D\right]^{\dagger}$ & - & 0.422 & \\
        $\chi_1$ & $1$ & $0.79$ & \\
        $\chi_2$ & - & $-3.32$ & \\
        \midrule
        $n_{env}$                       & 8                    & 4                    & No. of simulations per PPO iteration. \\
        $\alpha^{actor}$                  & \multicolumn{2}{c}{ \cref{fig:fig_learningrate} } & Actor learning rate. \\
        $\alpha^{entropy}$                & \multicolumn{2}{c}{ \cref{fig:fig_learningrate} } & Entropy penalty coefficient. \\
        $n_{epochs}^{actor}$            & \multicolumn{2}{c}{ 20                          } & Max. actor training epochs per iteration. \\
$\mathcal{T}_{ES}^{actor}$      & \multicolumn{2}{c}{ 0.025                       } & Early stopping threshold for actor.\\
        $\alpha^{critic}$                 & \multicolumn{2}{c}{ 0.001                       } & Critic learning rate. \\
        $n_{epochs}^{critic}$           & \multicolumn{2}{c}{ 50                          } & Max. critic training epochs per iteration. \\
${n}_{patience}^{critic}$       & \multicolumn{2}{c}{ 5                           } & Early stopping patience value for critic.\\
        $N_{PPO}$                & 500                  & 250                        & Max. PPO update iterations. \\
        $\delta$                 & \multicolumn{2}{c}{ 0.25                           } & State recycle fraction.\\
        \midrule
        $\text{auxvar}_{min}$ & 0.03 & 2.1 & \multirow{2}{3.6cm}{Auxiliary variable bounds (see \cref{sec:meth-multijet-rlnotes})} \\
        &&& \\
        $\text{auxvar}_{max}$ & 0.25 & 10 & \\
\midrule
        Neurons                         & \multicolumn{2}{c}{ 256                         } & \multirow{2}{3.6cm}{No. of neurons and layers in the MLP actor and critic networks.}  \\
        &&&\\
        Layers                          & \multicolumn{2}{c}{ 2                           } & \\
\bottomrule
        & & & \hfill \smaller{$^{\dagger}$Unactuated}
    \end{tabular}
    \caption{Hyperparameters and their values.}\label{tab:hyperparameters}
\end{table}

\section{Zero Net Mass Flow Rate Multi-Jets}
\label{sec:methodology-multijets}

Enforcing the net mass flow rate across the employed synthetic jets to be zero is a good way of ensuring that no excess momentum is being injected into the flow, and the effects observed are a result of pure flow-field changes induced by the AFC strategy.
It also has a benefit in the context of real-world applications, namely that it negates the need for an extra fluid reservoir to manage the net momentum excess/deficit.
As mentioned previously, in the 2-jet scenario it is straightforward to implement since one need only predict a single jet's intensity and set the other to be its opposite value \cite{rabault2019accelerating,rabault2019artificial,vignon2023recent,vinuesa2024perspectives,GARCIA2025109913,suarez2025active,KURZ2025106854}.
The traditional approach for more than 2 jets has been to use a mean centering strategy \cite{wang2022deep, GARCIA2025109913, tang2020robust}.
To our knowledge, it has however not been noted in literature that this has the pitfall of leading to a non-injective mapping from the network outputs to the implemented jet-intensities.
To highlight this drawback, this framework is introduced and analyzed in the following subsection, following which we propose an alternative strategy that has the additional advantage of superior cost scaling (\textit{cost} as defined in the following subsections).\\

\subsection{Traditional Mean-Centering Approach}
\label{sec:znmf_tradjet_theory}

As stated in \cref{sec:meth-rl-actions}, the $i^{th}$ jet's mass flow rate $Q_i$ is predicted as $Q_i = Q^{max} f_i(\boldsymbol{a})$, with $Q^{max}$ being the maximum mass flow rate, $f_i(\boldsymbol{a})$ the modulated action and $\boldsymbol{a}$ the output of the actor network.
The function $f_i$ ensures that $f_i(\boldsymbol{a}) \in [-1, 1]$ and $\sum_{i=1}^N f_i(\boldsymbol{a}) = 0$, thus maintaining $\sum_{i=1}^N Q_i = 0$ and $|Q_i| \leq Q^{max} \ \forall \ i \in \{1,\cdots, N\}$.
The actions vector $\boldsymbol{a} \in [a_{min}, a_{max}]^N$ is simply a concatenation of the actions corresponding to each individual jet.
As such, $\sum_{i=1}^N a_i \neq 0$ and to find the function $f_i(\cdot)$ consider
\begin{align}
    b_i &= a_i - \frac{\sum_{j=1}^N a_j}{N} = \frac{\sum_{j=1, {j \neq i}}^N (a_i - a_j)}{N} \label{eq:meth_znmf_tradjet_bidef}
\end{align}
Clearly, $\sum_{i=1}^N b_i = 0$, so it satisfies the zero net mass flow rate requirement. To find the bounds on this new entity in order to normalize it to the range $[-1, 1]$, note
\begin{align}
    b_{i,max} &=    \max \left[  b_i \right] \\
              &=    \max \left[ \frac{\sum_{j=1, {j \neq i}}^N (a_i - a_j)}{N} \right] \nonumber\\
              &\leq \frac{\sum_{j=1, {j \neq i}}^N \max \left(a_i - a_j\right)}{N}    \nonumber\\
              &\leq \frac{\sum_{j=1, {j \neq i}}^N \left(a_{max} - a_{min}\right)}{N}    \nonumber\\
    \implies b_{i,max} &\leq \left( \frac{N-1}{N} \right) \Delta a,
\end{align}
with $\Delta a = a_{max} - a_{min}$, and $b_{i,max}$ being the maximum attainable value for $b_i$.
This extremum for $b_i$ is then attained in the case $a_i=a_{max}$ and $a_j=a_{min} \forall j\neq i$, allowing the removal of the inequality
\begin{align}
    b_{i,max} = \left( \frac{N-1}{N} \right)  \Delta a  \end{align}
Similarly $b_{i,min} = -\left( \frac{N-1}{N} \right) \Delta a $.
Note that these bounds are independent of the index $i$.
Therefore, using these to scale all $b_i$, the modulating function $\boldsymbol{f}(\cdot)$ becomes
\begin{align}
    f_i(\boldsymbol{a}) &= \frac{N}{(N-1) \Delta a} b_i \nonumber\\
                    &= \frac{N}{(N-1) \Delta a} \left( a_i - \frac{\sum_{j=1}^N a_j}{N} \right), \label{eq:multijet-traditional}
\end{align}
transforming any $\boldsymbol{a} \in [a_{min}, a_{max}]^N$ into a normalized zero net mass flow rate compatible output (i.e. obeying $\sum_{i=1}^N f_i(\boldsymbol{a}) = 0$ and $-1 \leq f_i(\boldsymbol{a}) \leq 1$).
Now consider the \textit{running cost} of operating this momentum-transfer system, defined here to be
\begin{align}
    \mathcal{C} &= \sum_{i=1}^N |Q_i| \label{eq:cost-jetAFC} = Q^{max} \sum_{i=1}^N |f_i(\boldsymbol{a})|
\end{align}
Of course this quantity will vary with time throughout each simulation, so a good measure to is to check the maximum possible cost.
As detailed in \cref{sec:app-cost-oldapp}, for this case $\mathcal{C}_{max}$ turns out to be $\mathcal{C}_{max} = 2 \lfloor N/2\rfloor \left( \frac{N-\lfloor N/2\rfloor}{N-1}\right) Q^{max}$ ($\lfloor \cdot \rfloor$ being the floor function).
This exhibits a near linear scaling, scaling as $\sim N/2$ for large $N$. \\

The disadvantage of this approach is that owing to the mean-centering of the actions, the modulating function $\boldsymbol{f}(\cdot)$ produces a non-injective mapping from the actor outputs to the implemented jet-intensities.
This means that the modulating function introduced to ensure zero net-mass-flow-rate, could result in ambiguous implemented jet-actuations.
To elucidate, consider an action vector $\boldsymbol{a}_1$ resulting in certain modulated outputs $f_i (\boldsymbol{a}_1)$.
Now consider another distinct action vector $\boldsymbol{a}_2$ such that $a_{2, i} = a_{1, i} + c$, $c$ being a scalar constant (i.e. just an affine shift of the discrete distribution).
Then, according to \cref{eq:multijet-traditional},
\begin{align}
    f_i(\boldsymbol{a}_1) &= \frac{N}{(N-1) \Delta a} \left( a_{1,i} - \frac{\sum_{j=1}^N a_{1,j}}{N} \right) \nonumber
\end{align}
and,
\begin{align}
    f_i(\boldsymbol{a}_2) &= \frac{N}{(N-1) \Delta a} \left( a_{2,i} - \frac{\sum_{j=1}^N a_{2,j}}{N} \right) \nonumber \\
                      &= \frac{N}{(N-1) \Delta a} \left( (a_{1,i} + c) - \frac{\sum_{j=1}^N (a_{1,j}+c)}{N} \right) \nonumber \\
&= f_i(\boldsymbol{a}_1). \nonumber
\end{align}

Thus, linearly shifted actions become indistinguishable.
This clearly leaves open the possibility for the incorrect implementation the system's requested control ($\boldsymbol{a}$) by the modulating function ($\boldsymbol{f}(\boldsymbol{a})$), and undesirable outcome.
It can also hamper learning in a DRL context, as the agent would be unable to explore unique strategies since differing action outputs would be likely to produce close-to-similar jet intensities.
Logically following form this argument then, is the outcome that the agent settles on safely differentiable policies such as near constant suction/blowing, which is what has been observed in previous mean-centering multi-jet studies \cite{wang2022deep, GARCIA2025109913, tang2020robust}.
This hypothesis is tested and discussed up in \cref{sec:results}.

\subsection{Alternative Approach}
\label{sec:meth-multijet-altapproach}

Instead of the above approach, one can use a key insight from the standard 2-jet case, that is, only a single jet's intensity is predicted and the other automatically set to its negative value.
If only $N-1$ jet intensities are predicted, the $N^{th}$ jet's intensity can automatically be set as $f_N(\boldsymbol{a})= -\sum_{i=1}^{N-1} f_i(\boldsymbol{a})$ so that $\sum_{i=1}^N f_i(\boldsymbol{a}) = 0$.
In this case, $\boldsymbol{a} \in [a_{min}, a_{max}]^{N-1}$, and taking inspiration from multinomial logistic regression,
\begin{align}
    b_i = \begin{cases}
        \frac{a_i}{1 + \sum_{j=1}^{N-1} a_j} & \ \ \text{if} \ i \leq N-1, \\
        \frac{1}{1 + \sum_{j=1}^{N-1} a_j}   & \ \ \text{if} \ i = N
    \end{cases} \label{eq:meth-newapp-b_i}
\end{align}
Now $\sum_{i=1}^N b_i = 1$, and to ensure the denominator is never zero we set the condition $a_{min} = 0$.
The bounds on $b_i$ can be derived by enforcing $\boldsymbol{a} \in \mathbb{R}_{\geq 0}^{N-1}$, yielding $\max\left[b_i\right] = 1$ and $\min\left[b_i\right] = 0$.

To ensure a zero net mass flow rate (recall $\sum_{i=1}^N b_i = 1$), consider
\begin{align}
    \sum_{i=1}^N b_i' = \sum_{i=1}^N \left(b_i - \frac{1}{N}\right) = 0, \nonumber
\end{align}
resulting in $b_i' \in \left[\frac{-1}{N}, \frac{N-1}{N}\right]$. This leads to the final modulated actions being
\begin{align}
    f_i(\boldsymbol{a}) &= \left( \frac{N}{N-1} \right)  b_i' \nonumber \\
                    &= \begin{cases}
                        \left( \frac{N}{N-1} \right) \left( \frac{a_i}{1 + \sum_{j=1}^{N-1} a_j} - \frac{1}{N} \right)  & \ \ \text{if} \ i \leq N-1, \\
                        \left( \frac{N}{N-1} \right) \left( \frac{1}{1 + \sum_{j=1}^{N-1} a_j} - \frac{1}{N} \right)   & \ \ \text{if} \ i = N
                    \end{cases} \label{eq:meth-newapp-fi}
\end{align}
thereby avoiding the non-injectiveness of the previous approach, as shown below.

To prove a function $\mathbf{h}(\mathbf{x})$ is injective, one needs to prove $\mathbf{h}(\mathbf{x}_1) = \mathbf{h}(\mathbf{x}_2) \implies \mathbf{x}_1 = \mathbf{x}_2$.
For the function $\boldsymbol{f}(\boldsymbol{a})$, assume two action vectors $\boldsymbol{a}_1, \boldsymbol{a}_2 \in \mathbb{R}_{\geq0}^{N-1}$ such that $f_i(\boldsymbol{a}_1) = f_i(\boldsymbol{a}_2) \ \ \forall \ \ i \in \{1,\cdots,N\}$.
From \cref{eq:meth-newapp-fi}, for $i=N$, this gives
\begin{align}
    f_N(\boldsymbol{a}_1) &= f_N(\boldsymbol{a}_2) \nonumber \\
    \implies \sum_{j=1}^{N-1} a_{1,j} &= \sum_{j=1}^{N-1} a_{2,j}, \label{eq:app-inj-fN}
\end{align}
and applying this to $i < N$ leads to
\begin{align}
    f_i(\boldsymbol{a}_1) &= f_i(\boldsymbol{a}_2) \nonumber \\
    \implies \frac{a_{1,i}}{1 + \sum_{j=1}^{N-1} a_{1, j}} &= \frac{a_{2,i}}{1 + \sum_{j=1}^{N-1} a_{2,j}} \nonumber \\
    \implies a_{1,i} &= a_{2,i} .
\end{align}
Thus, under the new formulation $\boldsymbol{f}(\boldsymbol{a}_1) = \boldsymbol{f}(\boldsymbol{a}_2) \implies \boldsymbol{a}_1 = \boldsymbol{a}_2$, and it is indeed an injective mapping.

Note that $f_i(\boldsymbol{a}) \in \left[\frac{-1}{N-1}, 1\right] \subseteq [-1, 1]$, and therefore $|Q_i| = |Q^{max} f_i(\boldsymbol{a})|\leq Q^{max} \ \forall \ i \in \{1, \cdots, N\}$.
Additionally, the traditional 2-jet system emerges as a special case of this framework for $N=2$.

The non-symmetric bounds on $f_i$ introduce a kind of \textit{bias} towards higher possible positive values, while restricting the lowest possible negative values.
This leads to the possibility of another modulating function $\boldsymbol{g}(\cdot) = -\boldsymbol{f}(\cdot)$, that ensures both $\sum_{i=1}^N g_i = 0$ and $g_i \in \left[-1, \frac{1}{N-1}\right] \subseteq [-1, 1]$, now introducing an opposite bias via lower possible negative values and restricted positive values.
In this study, we call $\boldsymbol{g}(\cdot) = -\boldsymbol{f}(\cdot)$ the \textit{inverted} formulation and $\boldsymbol{f}(\cdot)$ the \textit{non-inverted}.
Both these modulating functions are implemented and explored in this study, with intriguing implications on the learned policies.

Lastly, the maximum cost of this system as derived in \cref{sec:app-cost-newapp} turns out to be $\mathcal{C}_{max} = 2 Q^{max}$, independent of the number of jets employed. This clearly shows that in addition to removing the non-injectiveness, this new formulation is also more cost-effective for higher jet counts ($N>3$).

\subsection{RL Specific Notes}
\label{sec:meth-multijet-rlnotes}

For the traditional mean-centering approach, as described in \cref{sec:znmf_tradjet_theory}, the actor network needs to predict $N$ $a_i$ values, with $a_i \in \left[ 0, 1 \right]$ in this study.
The modulating function $\boldsymbol{f}$ (from \cref{eq:multijet-traditional}) then subtracts the mean and scales them appropriately.
However, since in the proposed alternative approach $b_i \in [0, 1]$ and $a_i \geq 0$, instead of predicting $a_i$ we have the actor network predict $b_i$ values directly, ensuring that they have been entangled according to \cref{eq:meth-newapp-b_i}.

As mentioned in \cref{sec:methodology-rl}, the actor network in PPO is used to predict the defining parameters of a parameterized probability distribution.
The actions are then drawn from this distribution during the training/stochastic phase, and in the evaluation/deterministic phase the mode of the probability distribution is used as the action.
Hence, it is important to ensure that the modes of the used probability distribution follow the outlined framework.

In this study, two different distributions are used for the two different simulation setups.
For the cylinder setup, in order to maintain consistency with the previous study \cite{KURZ2025106854}, a Gaussian distribution is employed, defined as
\begin{align}
    p(x; \mu, \sigma) = \frac{1}{\sqrt{2\pi\sigma^2}} e^{-\frac{(x-\mu)^2}{2\sigma^2}}
\end{align}
where $p(x; \mu, \sigma)$ is the probability density function (PDF), $\mu$ is the mean and $\sigma$ the standard deviation.
Here the mode `$m$' and the mean $\mu$ turn out to be one and the same, so the actor network predicts the pairs $(\mu_i, \sigma_i)$ for sampling $a_i$ or $b_i$ (as required).
This setup ensures that in the evaluation/deterministic phase, \cref{eq:multijet-traditional} and \cref{eq:meth-newapp-b_i} and all the ensuing analyses are satisfied \textbf{exactly}.
However, the support for this distribution is $x \in (-\infty, \infty)$, meaning that even if the modes are confined to the interval $[0,1]$, the \textit{sampled} actions could theoretically be anywhere in the non-finite support.
Thus, during the training/stochastic phase, while this approach ensures the zero net mass flow rate requirement for the sampled actions \textbf{exactly} ($\sum_{i=1}^N Q_i = 0$), it could violate the condition $|Q_i| \leq Q^{max}$.
We have noticed our simulations crash during training at times, with the culprit being large $Q_i$ magnitudes far outside the allowed range.
This behavior can be limited by placing reasonable bounds on $\sigma_i$.
However, this is a soft constraint and does not guarantee $|Q_i|\leq Q^{max}$, only that the probabilities of such large predictions become minimal.

To deal with this non-finite support issue, in the airfoil's case a beta distribution is used to predict the actions,
\begin{align}
    p (x; \alpha, \beta) &= \frac{x^{\alpha-1} (1-x)^{\beta-1}}{B(\alpha, \beta)} \label{eq:meth_beta_dist}\ , \\
    B(\alpha, \beta) &= \frac{\Gamma(\alpha) \Gamma(\beta) }{\Gamma(\alpha + \beta)} \nonumber
\end{align}
with $p(x; \alpha, \beta)$ being the PDF as before, $\Gamma(\cdot)$ the gamma function and $\alpha, \beta$ the defining parameters.
Clearly, the support for this PDF is $x \in [0, 1]$, which aligns with the earlier requirements on $a_i$/$b_i$ values.
The mode of this distribution turns out to be
\begin{align}
    m = \frac{\alpha - 1}{\alpha + \beta - 2} \ \ \ \forall \ \alpha, \beta > 1.
\end{align}
Thus, to enforce the zero net mass flow rate framework, an alternative parameterization predicated on $m$ and $\nu=\alpha+\beta$ values instead of $\alpha, \beta$ is used.
The two are related as follows
\begin{align}
    \alpha &= m (\nu - 2) + 1 \\
    \beta  &= (1-m)\nu + 2m - 1.
\end{align}
The actor network then predicts the pairs $(m_i, \nu_i)$  (with $\nu_i > 2$) to define the individual beta distributions for sampling\footnotemark.
This means that one never observes out-of-bounds behavior for the mean-centering scheme.
In the case of the proposed scheme, in the evaluation/deterministic phase, this leads to \cref{eq:meth-newapp-b_i} and all ensuing analyses being satisfied \textbf{exactly}.
During the training/stochastic phase, this ensures that the \textit{sampled} actions satisfy the zero net mass flow rate requirement \textbf{exactly} ($\sum_{i=1}^N Q_i = 0$), while only altering bounds on the final sampled $f_N$ value to become $\left[-\left(N-\frac{1}{N-1}\right), 1\right]$.
This is better than the earlier Gaussian case as the outside-permissible-range behavior is limited to a single jet, and even then the range is bounded and not infinite.
\footnotetext{In practice, we set bounds on $\nu_i$, such that $\nu_i \in [\nu_{min}, \nu_{max}]$ with $2 < \nu_{min} < \nu_{max}$, to ensure the PDFs aren't too flattened/peaked (see \cref{tab:hyperparameters}).}

\section{Results}
\label{sec:results}

Both simulation setups are first advanced until $t^*=100$ with no actuation, at which point the initial transient behaviour is observed to have died out.
Statefiles taken from this point onward are used as initial states for the simulations carried out for training the DRL agent.
All the reported cases are then trained on three different random initializations for the DRL networks, to test the aforementioned training repeatability.
We use the following improvement metric to measure performance
\begin{align}
    \eta_{X} = \frac{\langle C_X\rangle^{AFC}}{\langle C_X\rangle^{unactuated}} - 1,
\end{align}
with $\langle C_X \rangle^{AFC}$ and $\langle C_X \rangle^{unactuated}$ are the time-averaged coefficients of the baseline case with no control and the case with AFC enabled.
It measures the change in the quantity $C_X$ with respect to the unactuated flow that the AFC is supposed to improve.

\subsection{Cylinder-in-Channel}

Here we keep the same jet-strength upper limit from \citet{KURZ2025106854}, $Q^{max} = 0.067 Q_{ref}$, where $Q_{ref}$ is defined to be the mass flow rate intercepting the cylinder.
Two jet position configurations are tested here, the first canonical and the second novel:\begin{itemize}
    \item (2 jets) $\varphi = \left\{ +90^{\circ}, -90^{\circ} \right\}$
    \item (4 jets) $\varphi = \left\{ +30^{\circ}, -30^{\circ}, +90^{\circ}, -90^{\circ} \right\}$
\end{itemize}
The angular width of the jets $\omega_i=10^{\circ}$ is the same for all jets in both configurations.
The canonical jet positioning is in line with earlier cited studies.
The novel configuration has been chosen so as to test the validity of the multi-jet frameworks for larger $N$ values, and the new positions picked to allow the agent the best chances at managing the separated region at the leeward side.

\subsubsection{Training Behavior}

\begin{figure}[t]
    \centering
    \tikzsetnextfilename{fig_cyl_2jet_rewards}
    \includegraphics{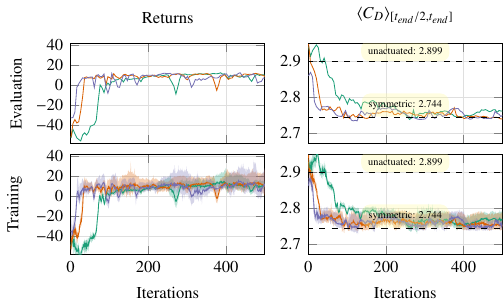}
    \caption{Training and evaluation metrics for the 2-jets cylinder system (across three random initializations).}
    \label{fig:rez_cyl_2jets}
\end{figure}

\begin{figure}[t]
    \centering
    \tikzsetnextfilename{fig_cyl_4jet_rewards-correctdirn}
    \includegraphics{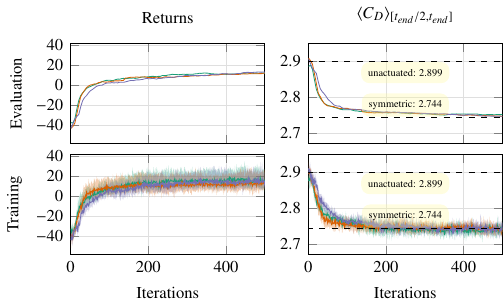}
    \caption{Training and evaluation metrics for the non-inverted 4-jets cylinder system (across three random initializations).}
    \label{fig:rez_cyl_4jets-correctdirn}
\end{figure}

\begin{figure}[t]
    \centering
    \tikzsetnextfilename{fig_cyl_4jet_rewards-fleppeddirn}
    \includegraphics{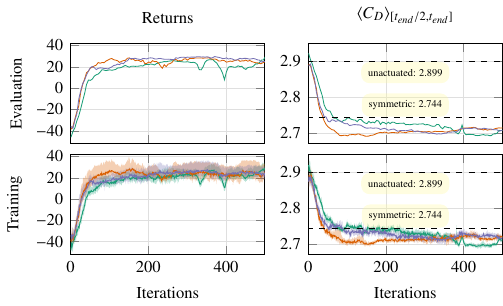}
    \caption{Training and evaluation metrics for the inverted 4-jets cylinder system (across three random initializations).}
    \label{fig:rez_cyl_4jets-fleppeddirn}
\end{figure}

\begin{figure}[t]
    \centering
    \tikzsetnextfilename{fig_cyl_4jet_rewards-meancentered}
    \includegraphics{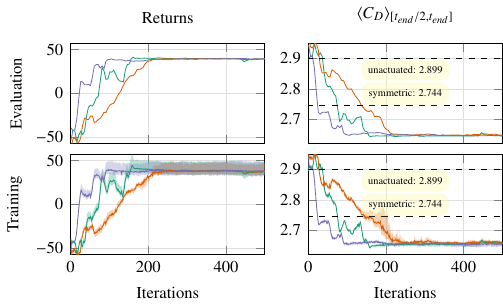}
    \caption{Training and evaluation metrics for the mean-centered 4-jets cylinder system (across three random initializations).}
    \label{fig:rez_cyl_4jets-meancentered}
\end{figure}

During training, we monitor the returns collected by the policy along with the average $C_D$ from the second half of the simulation, the latter being another indicator of policy performance.
Both these quantities are shown in \cref{fig:rez_cyl_2jets,fig:rez_cyl_4jets-correctdirn,fig:rez_cyl_4jets-fleppeddirn,fig:rez_cyl_4jets-meancentered}, each corresponding to the 2-jets, 4-jets non-inverted, inverted and mean-centered cases respectively.
In the training plots, the solid line marks the average values across the parallel simulations, whereas the shaded regions denote the spread between their respective minima and maxima.
As can be observed, learning is extremely stable and fast across the three different initializations for all cases.
This kind of repeatability indicates that performance is not a fluke, and the DRL agent is consistently capable of finding robust AFC policies.
Further note that the spread in training returns grows as the policy becomes better.
This is due to the implemented state-recycling between iterations.
As the policy improves, it leaves the flow in a higher-reward state than the unactuated flow.
A fraction $\delta$ of the initializing state files then comprises of these better starting points, and those trajectories naturally produce higher returns than those starting from the unactuated baseline states.
This leads to higher training returns for this fraction of the simulation cohort, thereby increasing the spread of the training metrics.

Just from these training curves themselves, we can observe that the 2-jets and 4-jets (non-inverted) cases collect similar returns and have similar last-half-average $C_D$ values.
The 4-jets (inverted) case collects much higher returns and has lower last-half-average $C_D$ values, with the 4-jets (mean-centered) collecting the highest and having the lowest last-half-average $C_D$ values.
Policies in both cases surpass the symmetric case also.
This behavior is further explored in the following subsection.

\subsubsection{Performance of Trained Policies}

\begin{table*}[]
    \addtolength{\tabcolsep}{-3pt}
    \begin{tabular*}{\linewidth}{@{\extracolsep{\fill}} lllllll}
\toprule & \textbf{2-Jets}   & \textbf{4-Jets}      & \textbf{4-Jets}   &  \textbf{4-Jets}     & \textbf{Unactuated}   & \textbf{Symmetric}     \\
        &                   & \textbf{(non-inverted)}  & \textbf{(inverted)}   &  \textbf{(mean-centered)} &          &                        \\
        \midrule $\langle{C_D}\rangle$             & $2.75\text{E}00$ & $2.74\text{E}00$  & $2.69\text{E}00$ & $2.64\text{E}00$       & $2.90\text{E}00$  & $2.74\text{E}00$ \\
        $r_D$                             & $-4.4 \%$        & $-0.8 \%$         & $+37.2 \%$       & $\boldsymbol{+68.1\%}$ & $--$              & $--$   \\
        $\eta_D$                          & $-5.0 \%$        & $-5.2 \%$         & $-7.1 \%$        & $\boldsymbol{-8.7\%}$  & $--$              & $--$   \\
$C_D^{\text{rms}}$                & $6.21\text{E-}03$ & $1.29\text{E-}02$ & $3.98\text{E-}03$       & $7.05\text{E-}03$ &  $1.99\text{E-}02$ & $0.0$  \\
        $\eta_D^{\text{rms}}$             & $-68.9 \%$        & $-35.4 \%$        & $\boldsymbol{-80.1 \%}$ & $-64.7\%$ &  $--$              & $--$   \\
        \midrule
        $\langle{C_L}\rangle$             & $6.61\text{E-}02$       & $-1.18\text{E-}02$ & $-1.51\text{E-}01$ & $-1.87\text{E-}02$ & $0.0$             & $0.0$ \\
$C_L^{\text{rms}}$                & $1.42\text{E-}01$       & $2.69\text{E-}01$  & $1.51\text{E-}01$  & $2.48\text{E-}01$ & $5.51\text{E-}01$ & $0.0$ \\
        $\eta_L^{\text{rms}}$             & $\boldsymbol{-74.3 \%}$ & $-51.1 \%$         & $-72.6 \%$         & $-54.9\%$ & $--$              & $--$   \\
        \midrule
        $\langle{C_{F}}\rangle$           & $2.75\text{E}00$  & $2.76\text{E}00$  & $2.70\text{E}00$ & $2.65\text{E}00$       & $2.95\text{E}00$  & $2.74\text{E}00$ \\
        $r_F$                             & $-5.5\%$          & $-7.1\%$          & $+23.5\%$        & $\boldsymbol{+44.9\%}$ & $--$              & $--$   \\
        $\eta_F$                          & $-6.5\%$          & $-6.4\%$          & $-8.5\%$         & $\boldsymbol{-9.9\%}$  & $--$              & $--$   \\
$C_F^{\text{rms}}$                & $6.80\text{E-}03$       & $2.10\text{E-}02$ & $8.27\text{E-}03$ & $1.45\text{E-}02$ & $5.67\text{E-}02$ & $0.0$  \\
        $\eta_F^{\text{rms}}$             & $\boldsymbol{-87.9 \%}$ & $-62.5 \%$        & $-85.2 \%$        & $-74.2\%$ & $--$              & $--$   \\
        \midrule
        $\langle C^* \rangle$                             & $\boldsymbol{3.08\text{E-}02}$ & $1.10\text{E-}01$ & $1.01\text{E-}01$ & $1.76\text{E-}01$ & $--$ & $--$ \\
        \bottomrule \end{tabular*}
    \addtolength{\tabcolsep}{3pt}
    \caption{Time-averaged force coefficients $\langle{C_X}\rangle$ as well as their root-mean-squared values $C_X^{\text{rms}}$ for the cylinder-in-channel setup, computed and averaged over a time period $t^*\in[50,100]$ well after the training time in the quasi-stable limit of the controlled flow. $C^*$ is the normalized running cost, defined to be $C^* = \sum_{i=1}^N | Q^*_i |$, with $Q^*_i$ being the normalized jet-strengths.}\label{tab:rez_cyl_clcdcf_latterhalf_extendtime}
\end{table*}

\begin{figure}[t]
    \centering
    \tikzsetnextfilename{fig_cyl_clcd-extended_time}
    \includegraphics{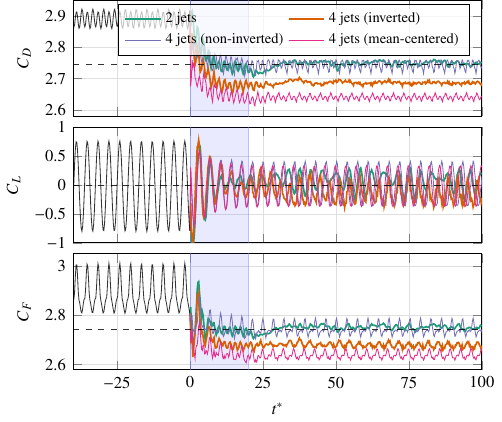}
    \caption{Long-term evolution of the lift, drag and total force coefficients with the AFC agents enabled on the cylinder. Control starts at $t^*=0$, while the shaded interval is the simulation time used for training the agent. $t^*<0$ is the unactuated flow, and the dashed line (- - -) represents the symmetric case.}
    \label{fig:rez_cyl_clcd}
\end{figure}

\begin{figure}[t]
    \centering
    \tikzsetnextfilename{fig_cyl_action-extended_time}
    \includegraphics{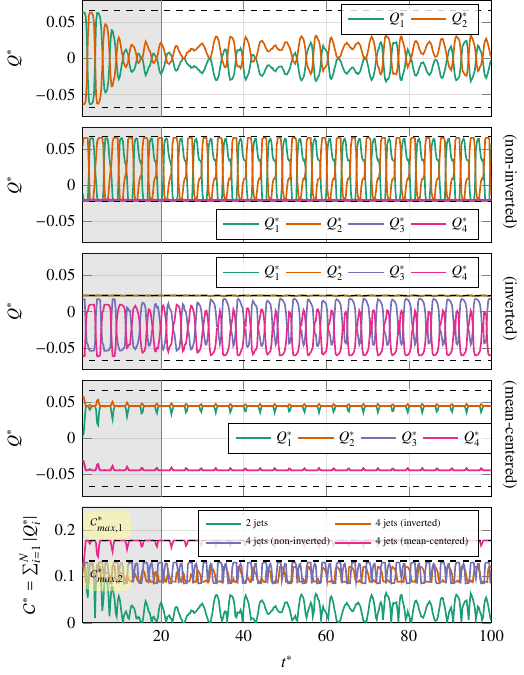}
    \caption{Jet strengths and their associated costs for the jet setups on the cylinder. $Q^*_i$ values are the jet-strengths normalized by $Q_{ref}$, it being the mass flow rate intercepted by the cylinder. In the 2-jets case, $Q_1^*$ and $Q_2^*$ correspond to $\varphi_i = \{90^{\circ}, -90^{\circ}\}$. In the 4-jets case, $Q_1^*$, $Q_2^*$, $Q_3^*$ and $Q_4^*$ correspond to jets located at $\varphi_i = \{30^{\circ}, -30^{\circ}, 90^{\circ}, -90^{\circ}\}$. The bounds on the jet strengths (\cref{eq:meth-newapp-fi}) are shown by the dashed lines (- - -) in the $Q^*$ plots. $\mathcal{C}^*$ is the non-dimensionalized cost, computed as described, where the dashed lines represent the maximum possible costs. $\mathcal{C}^*_{max,1}$ is the max. cost for the traditional mean-centered approach and $\mathcal{C}^*_{max,2}$ for the proposed approach. Clearly, all 4-jets cases operate at or close to their respective maximum possible costs.}
    \label{fig:rez_cyl_action}
\end{figure}

Since we have an idealized symmetric case here, we use an additional \textit{relative} improvement metric from \citet{KURZ2025106854}
\begin{align}
    r_{X} = \frac{\langle C_X\rangle^{unactuated} - \langle C_X\rangle^{AFC}}{\langle C_X\rangle^{unactuated} - \langle C_X\rangle^{sym}} - 1
\end{align}
where $\langle C_X\rangle^{sym}$ is the time-averaged coefficients of this symmetric case.
$r_X = 0$ indicates the AFC case achieves the same reduction in $C_X$ as the symmetric case and $r_X > 0$ denotes that the AFC case is able to suppress $C_X$ below that of the symmetric case.
The symmetric case has no vortex shedding, which can be thought of as a good estimate of the maximum drag reduction achievable through AFC.

The best performing policies are taken from across the different initializations and used here for plotting and discussion.
The cases while only having been trained on simulations with $t_{end}^*=20$, are evaluated on simulations lasting up to $t^*=100$.
The resulting $C_L$, $C_D$ and $C_F$ evolution for the three cases are shown in \cref{fig:rez_cyl_clcd} along with the jet strengths and running costs in \cref{fig:rez_cyl_action}.
While the observation of the 2-jets and 4-jets (non-inverted) systems performing similarly from the previous subsection is true in an aggregate sense, the 4-jets (non-inverted) system offers much more stable $C_L$, $C_D$ development.
The 2-jets case leads to a rapid reduction, followed by an increase and settlement into a quasi-stable (but still fluctuating) behavior.
The 4-jets (non-inverted) case on the other hand follows a smooth descent settling into a steadily periodic pattern.
The inverted case offers similar lift damping, but superior drag and total force reductions compared to these both, while the mean-centered case leads to the maximum drag and total force reductions but modest lift damping.

The averaged coefficients from the latter half of the simulations are reported in \cref{tab:rez_cyl_clcdcf_latterhalf_extendtime}, along with their root-mean-squared values.
The RMS values $C_X^{\text{rms}}$ are defined to be
\begin{align}
    C_X^{\text{rms}} = \sqrt{ \langle \ \left( \ C_X - \langle C_X \rangle \ \right)^2 \ \rangle } \ \ ,
\end{align}
and are an indication of the spread around the mean in their respective quantities.
The numbers agree with the previous qualitative comments, in that the 2-jets and 4-jets (non-inverted) systems perform similarly in terms of their averages, with a roughly $5\%$ drag reduction and $6.5\%$ total force reduction.
However, the 4-jets (non-inverted) system has higher RMS values, indicating stronger oscillations which are also observable in \cref{fig:rez_cyl_clcd}.
Note that even though the oscillations are greater than the 2-jets case, they are still damped compared to the unactuated case ($\sim51\%$ lift-RMS reduction compared to $\sim74\%$ for 2-jets, for example).
The 4-jets (inverted) case offers the more impressive performance and comparable RMS values to the 2-jets case as well, with a $7.1\%$ drag reduction and $8.5\%$ total force reduction.
This results in $+37.2\%$ $r_D$ and $+23.5\%$ $r_F$ values, indicating that it is very successful in reducing drag and total force to beyond the symmetric case's levels too.
The mean-centered approach has the most drag and total-force reduction, with $\eta_D=8.7\%, \eta_F=9.9\%$, and $r_D=+68.1\%, r_F=+44.9\%$.
These $r_D,r_F$ values indicate that this too reduces drag and total-force to beyond the symmetric case's levels, even more so than the previous inverted case.
The lift damping is comparable to the non-inverted case at $\sim 55\%$, but not as much as the 2-jets and 4-jets (inverted) cases.
The 4-jets cases have a higher running cost than the 2-jets case, with the mean-centered approach having the highest.

Taking a look at \cref{fig:rez_cyl_action}, the standard 2-jets case settles upon the well-known $C_L$ opposed vorticity injection strategy in order to manage the wake and reduce lift and drag.
All 4-jets systems operate at or close to their respective maximum possible costs.
In all three of these cases, the agent learns policies that prefer negative/lower values for the $\varphi_i=\pm90^{\circ}$ jets and positive/higher values for the $\varphi_i=\pm 30^{\circ}$ jets.
This indicates a common strategic playbook despite the differing dynamics, where the agent finds ways to exploit the jet positions by using the $\varphi_i=\pm30^{\circ}$ to add a propulsive element.

In the mean-centered case, this is done in a straightforward way by constantly ingesting momentum at the poles and redirecting it to the $\pm 30^{\circ}$ jets for propulsion.
The 4-jets (inverted and non-inverted) systems find specialized AFC strategies that work well within their respective bounds.
In addition to the propulsive element at the leeward side, they also try to manage the wake by countering vortex shedding using periodic behavior.
The non-inverted case does this by constantly ingesting momentum at the $\varphi_i=\pm 90^{\circ}$ jets, and expelling it in a periodic fashion at the $\varphi_i=\pm 30^{\circ}$ jets.
The inverted case does this by constantly expelling momentum at the $\varphi_i=\pm 30^{\circ}$ jets while ingesting it at the poles periodically.
In both cases, periodic behavior is observed with a bias towards the higher magnitude bound, i.e. positive in the non-inverted case and negative in the inverted, yet in a manner consistent with maintaining the propulsive effect.
The second strategy allows for more effective wake control as well, since material exchange at the poles seems better suited at countering the shed vortices (as in the 2-jets case).
This fact is also apparent in the force evolution plots in \cref{fig:rez_cyl_clcd} and RMS values in \cref{tab:rez_cyl_clcdcf_latterhalf_extendtime}, where the 2-jets and 4-jets (inverted) cases are able to control these vortex-shedding induced fluctuations much more effectively.

\begin{figure}[tb]
    \centering
    \tikzsetnextfilename{fig_cyl_cp}
    \includegraphics{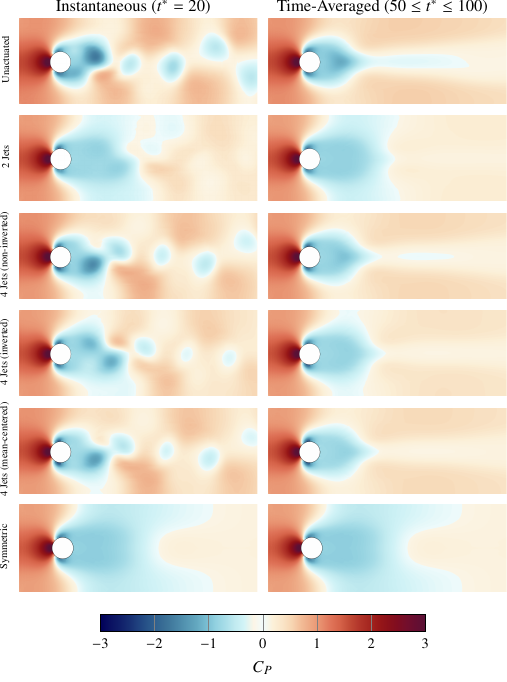}
    \caption{The pressure distribution throughout the domain, for the different flow-control cases in the cylinder-in-channel simulations. Domain clipped for visualization.}
    \label{fig:rez_cyl_cp}
\end{figure}

\begin{figure}[tb]
    \centering
    \tikzsetnextfilename{fig_cyl_velmag}
    \includegraphics{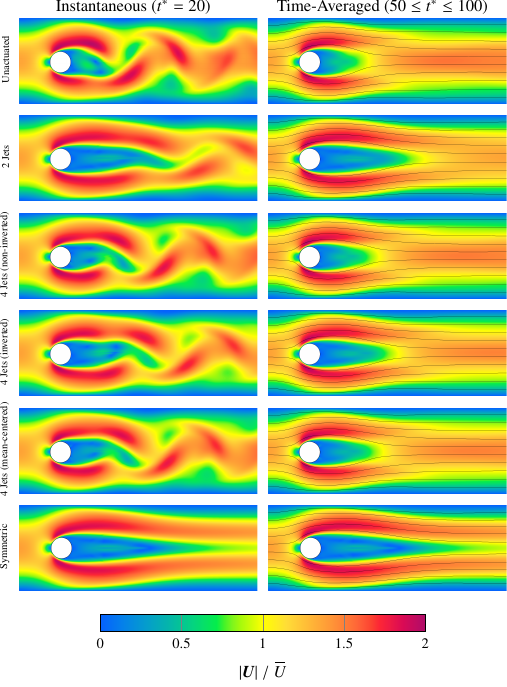}
    \caption{The velocity magnitude distribution throughout the domain, for the different flow-control cases in the cylinder-in-channel simulations. Streamlines have been added in the time-averaged cases to aid identification of the re-circulation region behind the cylinder. Domain clipped for visualization.}
    \label{fig:rez_cyl_velmag}
\end{figure}

\begin{figure}[tb]
    \centering
    \tikzsetnextfilename{fig_cyl_vortz}
    \includegraphics{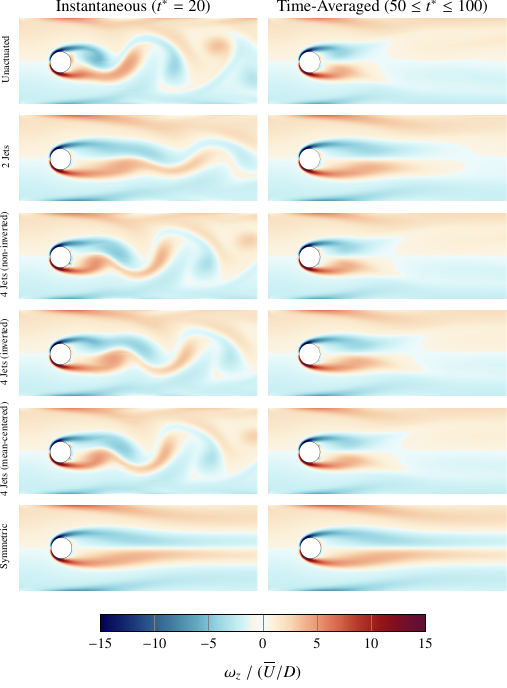}
    \caption{The $z$-direction vorticity distribution throughout the domain $\left( \omega_z = \frac{\partial U_y}{\partial x} - \frac{\partial U_x}{\partial y} \right)$, for the different flow-control cases in the cylinder-in-channel simulations. Domain clipped for visualization.}
    \label{fig:rez_cyl_vortz}
\end{figure}

The flow-fields are shown in \cref{fig:rez_cyl_cp,fig:rez_cyl_vortz,fig:rez_cyl_velmag}, and reflect the previous observations.
The 2-jets case is extraordinarily effective at controlling the wake, with the averaged flow-fields looking remarkably like the symmetric case and explaining its extreme fluctuation damping.
This is also reflected in its $r_D$ and $r_F$ values (\cref{tab:rez_cyl_clcdcf_latterhalf_extendtime}), both being close to zero.
The 4-jets cases on the other hand look more like the unactuated case but with lesser intensity vortices and a \textit{pushed back} wake, further demonstrating the propulsive element plus vortex management strategy.
The earlier observation of pole-side material exchange being more effective at wake control is apparent in the inverted case's plots, wherein the separation region is more elongated and the wake lesser in intensity (lower $C_P$ magnitudes) than the other 4-jets cases.

\subsection{Airfoil-in-Channel}

In the airfoil-in-channel setup, we follow the jet-strength limits from \citet{wang2022deep}, i.e. \mbox{$\sum_{i=1}^3 |Q_i| \leq 0.072 Q_{ref}$}, where $Q_{ref} = {\rho_{\infty}} \overbar{U} D$ is defined to be the reference mass flow rate.
In the proposed alternative approach, the quantity $\sum_{i=1}^N |Q_i|$ is known to be $\sum_{i=1}^N |Q_i| \leq 2 Q^{max}$.
This results in the upper limit $Q^{max} = 0.036 {\rho_{\infty}} \overbar{U} D$.
Two jet position configurations, a canonical and a novel case, are tested in this study:\begin{itemize}
    \item (3 jets) $\hat{x}_i / D = \left\{0.2, 0.3, 0.4 \right\}$ on the suction surface
    \item (6 jets) $\hat{x}_i / D = \left\{0.091, 0.491, 0.891 \right\}$ on both the suction and pressure surfaces
\end{itemize}
The jet width $\Delta_i/D=0.018$ is the same for all jets in both configurations.
As before, the canonical jet positioning is in line with previous studies \cite{wang2022deep,GARCIA2025109913}.
However, the major characterizing flow phenomena in this setup are the separated shear layer at the leading edge and the vortices shed from the trailing edge.
This could make it harder to control these phenomena by placing the jets far away from them, as in the first case.
Keeping this in mind, the second configuration is chosen to place actuators around these positions, and test the multi-jet frameworks on high $N$ values.
The same $Q^{max}$ values as computed above are used for the mean-centered cases as well, even when it results in a higher maximum possible cost ($N>3$).

\subsubsection{Training Behavior}

\begin{figure}[]
    \centering
    \tikzsetnextfilename{fig_af_3jet_tradplacement_Ma04_rewards}
    \includegraphics{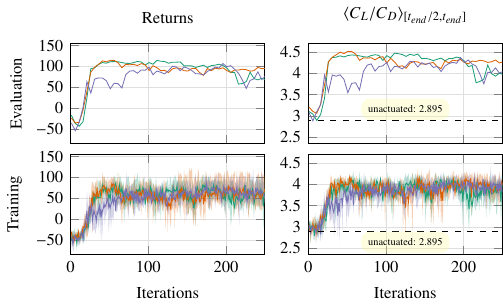}
    \caption{Training and evaluation metrics for the non-inverted 3-jets airfoil system (across three random initializations).}
    \label{fig:af_3jet_tradplacement_Ma04_rewards}
\end{figure}

\begin{figure}[]
    \centering
    \tikzsetnextfilename{fig_af_3jet_tradplacement_Ma04_rewards_invertedjet}
    \includegraphics{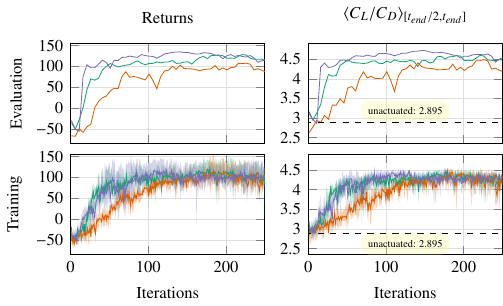}
    \caption{Training and evaluation metrics for the inverted 3-jets airfoil system (across three random initializations).}
    \label{fig:af_3jet_tradplacement_Ma04_rewards_invertedjet}
\end{figure}

\begin{figure}[tb]
    \centering
    \tikzsetnextfilename{fig_af_3jet_tradplacement_Ma04_rewards-meancentering}
    \includegraphics{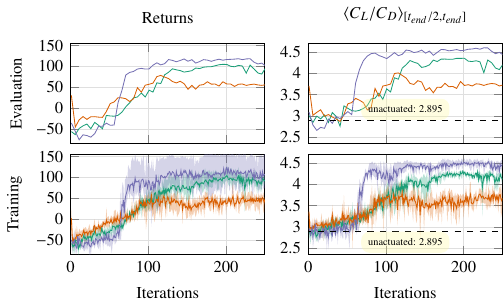}
    \caption{Training and evaluation metrics for the mean-centered 3-jets airfoil system (across three random initializations).}
    \label{fig:af_3jet_tradplacement_Ma04_rewards-meancentering}
\end{figure}

\begin{figure}[tb]
    \centering
    \tikzsetnextfilename{fig_af_6jets_Ma04_rewards}
    \includegraphics{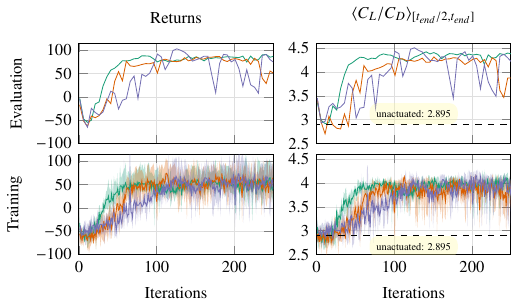}
    \caption{Training and evaluation metrics for the non-inverted 6-jets airfoil system (across three random initializations).}
    \label{fig:af_6jets_Ma04_rewards}
\end{figure}

\begin{figure}[tb]
    \centering
    \tikzsetnextfilename{fig_af_6jets_Ma04_rewards_invertedjet}
    \includegraphics{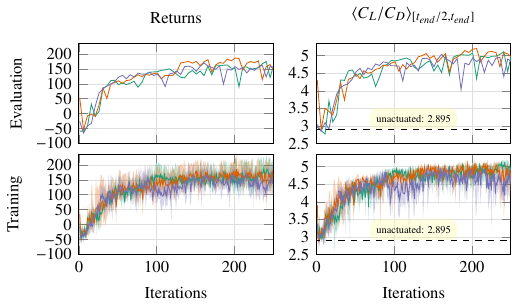}
    \caption{Training and evaluation metrics for the inverted 6-jets airfoil system (across three random initializations).}
    \label{fig:af_6jets_Ma04_rewards_invertedjet}
\end{figure}

\begin{figure}[tb]
    \centering
    \tikzsetnextfilename{fig_af_6jet_Ma04_rewards-meancentering}
    \includegraphics{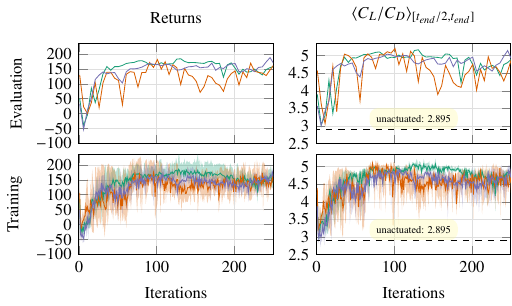}
    \caption{Training and evaluation metrics for the mean-centered 6-jets airfoil system (across three random initializations).}
    \label{fig:af_6jet_Ma04_rewards-meancentering}
\end{figure}

Since the aim here is to increase aerodynamic efficiency, instead of the drag we monitor the average of $C_L/C_D$ from the latter half of the simulation as an additional performance metric during training.
This and the collected returns are shown in \cref{fig:af_3jet_tradplacement_Ma04_rewards,fig:af_3jet_tradplacement_Ma04_rewards_invertedjet,fig:af_3jet_tradplacement_Ma04_rewards-meancentering,fig:af_6jets_Ma04_rewards,fig:af_6jets_Ma04_rewards_invertedjet,fig:af_6jet_Ma04_rewards-meancentering} for the 3-jets (non-inverted, inverted and mean-centered) and 6-jets (non-inverted, inverted and mean-centered) cases respectively.
Once again, in the training plots, the solid line marks the average values across the parallel simulations, whereas the shaded regions denote the spread between their respective minima and maxima.
Learning is quite stable for the proposed alternative scheme with consistent performance across random seeds, indicating methodological robustness.
Since the policies were observed to learn and converge quickly, training was limited to $250$ iterations (as opposed to $500$ in the cylinder setups).
However, for the 3-jets (mean-centering) cases, learning appears more initialization dependent, with different seeds leading to differing returns and last-half-average $C_L/C_D$ values (both in evaluation and training modes).
This is starkly visibly in \cref{fig:af_3jet_tradplacement_Ma04_rewards-meancentering}, and suggests that in a more complex scenario, the agents have a harder time learning with the mean-centered approach.
As noted previously, the spread in training metrics increases as the policy improves, owing to a fraction of the environments starting from improved states.

From these plots, it can be seen that the inverted configurations result in much higher returns than their non-inverted counter-parts for both 3 and 6 jets.
Additionally and quite interestingly, both the non-inverted configurations (3 and 6 jets) collect similar returns and have similar last-half-average $C_L/C_D$ values.
The 3-jets (inverted) case has slightly higher returns, while the 6-jets (inverted) case has the highest, with returns crossing $180$.
The best performing mean-centered 3 and 6 jets cases collect similar rewards as their respective inverted counterparts.
The last-half-average $C_L/C_D$ also show similar trends, with the inverted and mean-centered 6-jets cases showing the highest achieved aerodynamic efficiencies ($C_L/C_D$ values crossing $5$).

Lastly, note that despite the reduced informational awareness due to the conservative sensor placement, the agents learn effectively and without issues.
This lends credence to our initial hypothesis that in a subsonic context such as this one, surface sensors can indeed be enough for effective flow control.
It further showcases how careful sensor placement can help cut down on input size and remove redundant information, while still enabling efficient learning.

\subsubsection{Performance of Trained Policies}

\begin{table*}[]
    \addtolength{\tabcolsep}{-3pt}
    \begin{tabular*}{\linewidth}{@{\extracolsep{\fill}} llllllll}
        \toprule & \textbf{3-Jets}     & \textbf{3-Jets} & \textbf{3-Jets}        & \textbf{6-Jets}      & \textbf{6-Jets}   & \textbf{6-Jets}        &  \textbf{Unact.}   \\
        & \textbf{(non-inv.)} & \textbf{(inv.)} & \textbf{(mean-cent.)} & \textbf{(non-inv.)}  & \textbf{(inv.)}   & \textbf{(mean-cent.)} &                    \\
        \midrule $\langle{C_D}\rangle$             & $3.35\text{E-}01$ & $2.89\text{E-}01$ & $3.19\text{E-}01$ & $3.21\text{E-}01$ & $2.83\text{E-}01$      & $3.02\text{E-}01$ &  $3.28\text{E-}01$\\
        $\eta_D$                          & $+2.1 \%$         & $-11.8\%$         & $-2.9 \%$         & $-2.1\%$          & $\boldsymbol{-13.6\%}$ & $-7.9\%$          & $--$ \\
$C_D^{\text{rms}}$                & $5.57\text{E-}02$ & $5.06\text{E-}02$ & $7.05\text{E-}02$ & $1.79\text{E-}02$      & $2.08\text{E-}02$ & $5.72\text{E-}02$ & $4.41\text{E-}02$ \\
        $\eta_D^{\text{rms}}$             & $+26.3 \%$        & $+14.9\%$         & $+60.0 \%$        & $\boldsymbol{-59.4\%}$ & $-52.9\%$         & $+29.9\%$         & $--$ \\
        \midrule
        $\langle{C_L}\rangle$             & $1.50\text{E}00$       & $1.37\text{E}00$ & $1.42\text{E}00$ &  $1.43\text{E}00$ & $1.44\text{E}00$ & $1.50\text{E}00$ & $9.69\text{E-}01$ \\
        $\eta_L$                          & $\boldsymbol{+54.9\%}$ & $+41.0\%$        & $+46.8\%$        &  $+47.6\%$        & $+48.8\%$        & $+54.8\%$        & $--$ \\
$C_L^{\text{rms}}$                & $2.15\text{E-}01$ & $1.88\text{E-}01$ & $2.34\text{E-}01$ & $1.39\text{E-}01$ & $1.36\text{E-}01$      & $2.06\text{E-}01$ & $2.05\text{E-}01$ \\
        $\eta_L^{\text{rms}}$             & $+4.5\%$          & $-8.4\%$          & $+14.2 \%$        & $-32.5\%$         & $\boldsymbol{-33.8\%}$ & $+0.3\%$ & $--$ \\
        \midrule
        $\langle C_L/C_D \rangle$         & $4.50\text{E}00$  & $4.76\text{E}00$  & $4.54\text{E}00$ & $4.46\text{E}00$  & $5.10\text{E}00$       & $5.04\text{E}00$ & $2.94\text{E}00$ \\
        $\eta_{L/D} $                     & $+53.1\%$         & $+61.8\%$         & $+54.4 \%$       & $+51.7\%$         & $\boldsymbol{+73.6\%}$ & $+71.4\%$        & $--$\\
$(C_L/C_D)^{\text{rms}}$          & $1.74\text{E-}01$      & $2.25\text{E-}01$ & $3.85\text{E-}01$ & $4.26\text{E-}01$ & $4.85\text{E-}01$ & $4.59\text{E-}01$ & $4.13\text{E-}01$\\
        $\eta_{L/D}^{\text{rms}}$         & $\boldsymbol{-57.8\%}$ & $-45.5\%$         & $-6.72 \%$        & $+3.1\%$          & $+17.6\%$         & $+11.1\%$         & $--$\\
        \midrule
        $\langle C^* \rangle$                             & $4.76\text{E-}02$      & $6.35\text{E-}02$ & $3.54\text{E-}02$ & $\boldsymbol{3.02\text{E-}02}$ & $3.76\text{E-}02$ & $4.56\text{E-}02$ & $--$ \\
        \bottomrule \end{tabular*}
    \addtolength{\tabcolsep}{3pt}
    \caption{Time-averaged force coefficients $\langle{C_X}\rangle$ as well as their root-mean-squared values $C_X^{\text{rms}}$ for the airfoil-in-channel setup, computed and averaged over a time period $t^*\in[25,50]$ well after the training time in the quasi-stable limit of the controlled flow. $C^*$ is the normalized running cost, defined to be $C^* = \sum_{i=1}^N | Q^*_i |$, with $Q^*_i$ being the normalized jet-strengths.}\label{tab:rez_af_clcdcf_latterhalf_extendtime}
\end{table*}

\begin{figure}[h]
    \centering
    \tikzsetnextfilename{fig_af_clcd-extended_time}
    \includegraphics{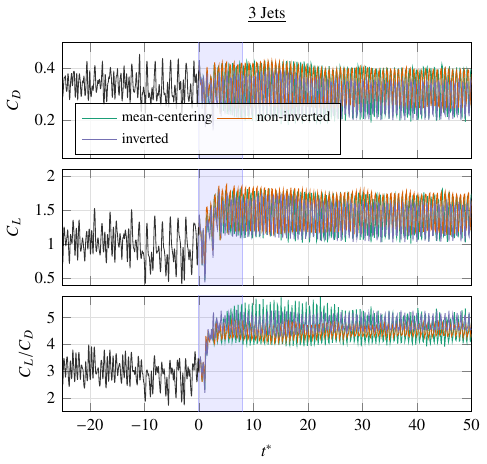}
    \caption{Long-term evolution of the lift, drag and aerodynamic efficiency coefficients with the AFC agents enabled, for the 3-jets cases in the airfoil-in-channel setup. Control starts at $t^*=0$, while the shaded interval is the simulation time used for training the agent. $t^*<0$ is the unactuated flow.}
    \label{fig:rez_af_clcd}
\end{figure}

\begin{figure}[tb]
    \centering
    \tikzsetnextfilename{fig_af_clcd-extended_time-6jets}
    \includegraphics{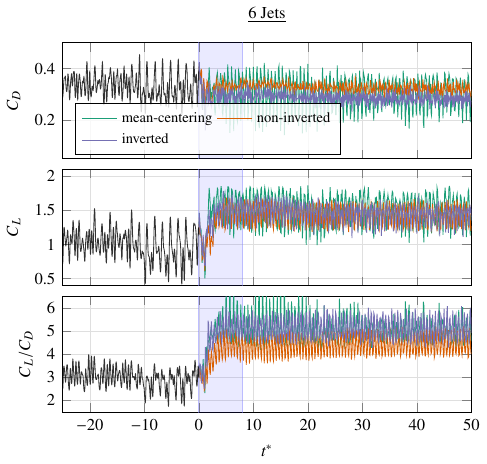}
    \caption{Long-term evolution of the lift, drag and aerodynamic efficiency coefficients with the AFC agents enabled, for the 6-jets cases in the airfoil-in-channel setup. Control starts at $t^*=0$, while the shaded interval is the simulation time used for training the agent. $t^*<0$ is the unactuated flow.}
    \label{fig:rez_af_clcd-6jets}
\end{figure}

\begin{figure}[t]
    \centering
    \tikzsetnextfilename{fig_af_3jet_action-extended_time}
    \includegraphics{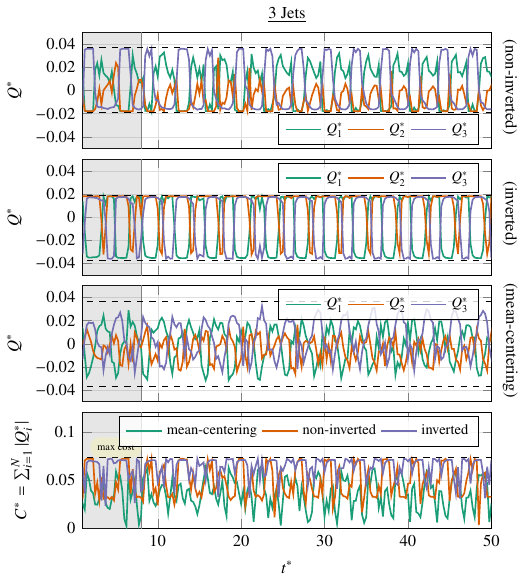}
    \caption{Jet strengths and their associated costs for the 3-jet setup on the airfoil. $Q^*_i$ values are the jet-strengths normalized by $Q_{ref}=\rho_{\infty}\overbar{U}D$. $Q_1^*$, $Q_2^*$ and $Q_3^*$ correspond to jets located at $\hat{x}_i/D = \{0.2, 0.3, 0.4\}$ on the suction surface of the airfoil. The bounds on the jet strengths (\cref{eq:meth-newapp-fi}) are shown by the dashed lines (- - -) in the $Q^*$ plots, while in the cost plot it represents the maximum cost (it being the same for both approaches for $N=3$).}
    \label{fig:rez_af_3jet_action_cost}
\end{figure}

\begin{figure}[]
    \centering
    \tikzsetnextfilename{fig_af_6jet_action-extended_time}
    \includegraphics{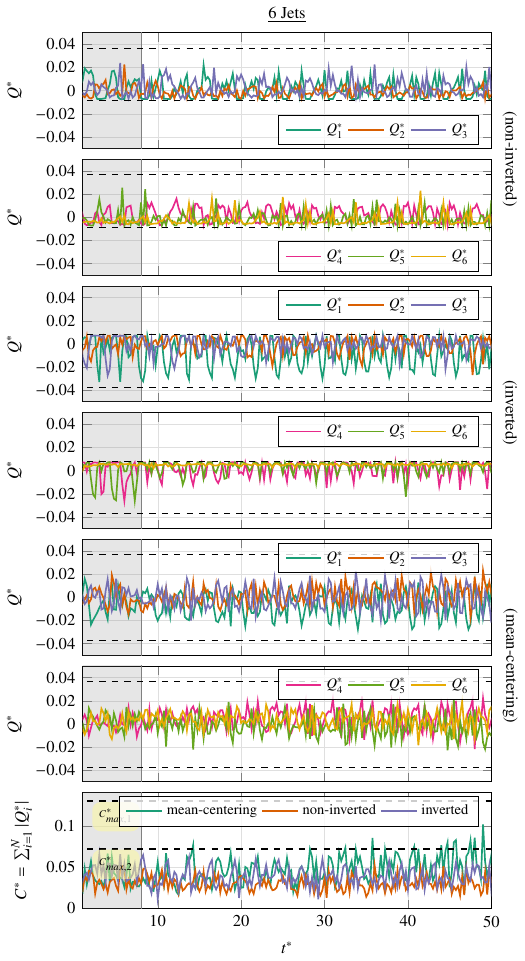}
    \caption{Jet strengths and their associated costs for the 6-jet setup on the airfoil. $Q^*_i$ values are the jet-strengths normalized by $Q_{ref}=\rho_{\infty}\overbar{U}D$. $Q_1^*$, $Q_2^*$, $Q_3^*$, $Q_4^*$, $Q_5^*$ and $Q_6^*$ correspond to jets located at $\hat{x}_i/D = \{0.091, 0.491, 0.891\}$ on the suction and pressure surfaces of the airfoil respectively. The bounds on the jet strengths (\cref{eq:meth-newapp-fi}) are shown by the dashed lines (- - -) in the $Q^*$ plots. In the cost plot it represents the maximum possible cost, $C^*_{max,1}$ being the max. cost for the traditional mean-centered approach and $C^*_{max,2}$ for the proposed approach. The mean-centering case can be seen breaching the proposed approach's cost ceiling at times, reflected in its higher averaged running cost as reported in \cref{tab:rez_af_clcdcf_latterhalf_extendtime}.}
    \label{fig:rez_af_6jet_action_cost}
\end{figure}

\begin{figure}[tb]
    \centering
    \tikzsetnextfilename{fig_af_3jet_tradplacement_Ma04_cp}
    \includegraphics{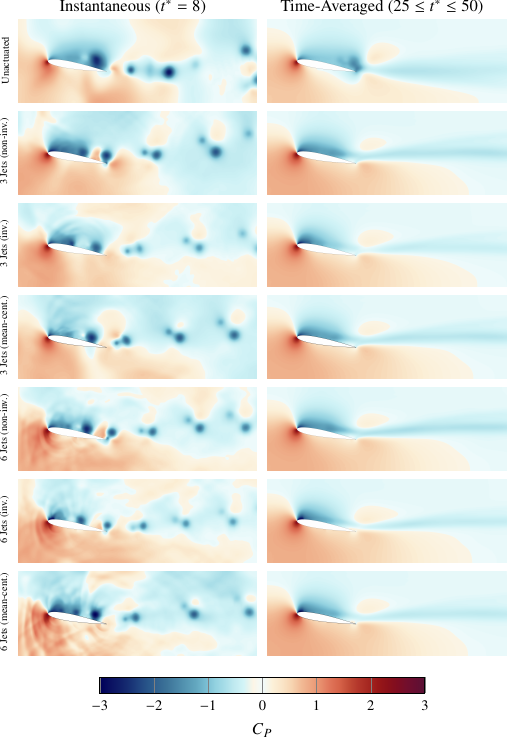}
    \caption{The pressure distribution throughout the domain, for the different flow-control cases in the airfoil-in-channel simulations. Interestingly, the acoustic waves emanating from the the jets are also visible in the instantaneous pressure plots. Domain clipped to exclude the sponge section.}
    \label{fig:rez_af_cp}
\end{figure}

\begin{figure}[tb]
    \centering
    \tikzsetnextfilename{fig_af_3jet_tradplacement_Ma04_velmag}
    \includegraphics{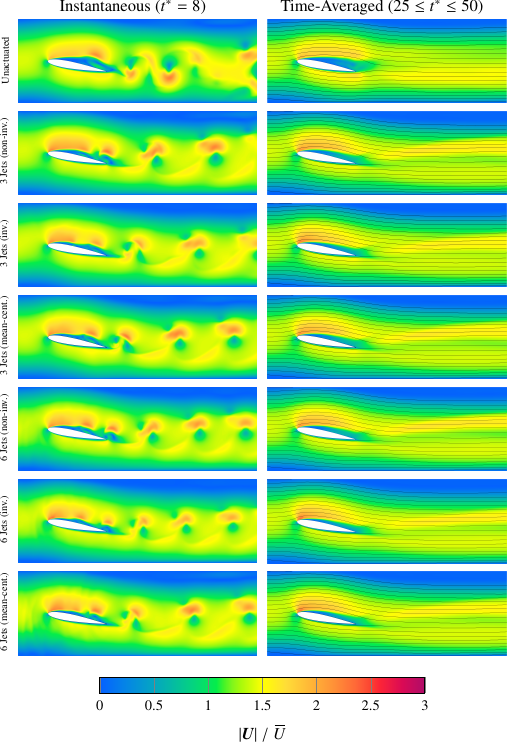}
    \caption{The velocity magnitude distribution throughout the domain, for the different flow-control cases in the airfoil-in-channel simulations. Streamlines added in the time-averaged cases for better visualization of the separated region. Domain clipped to exclude the sponge section.}
    \label{fig:rez_af_velmag}
\end{figure}

\begin{figure}[tb]
    \centering
    \tikzsetnextfilename{fig_af_3jet_tradplacement_Ma04_vortz}
    \includegraphics{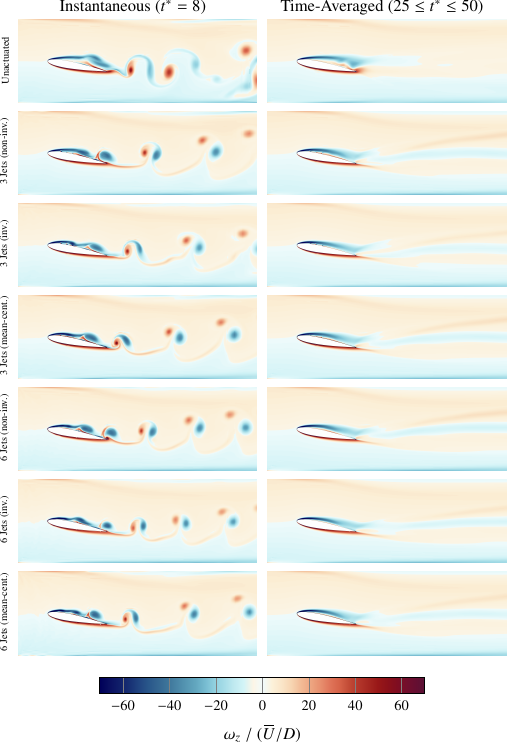}
    \caption{The $z$-direction vorticity distribution throughout the domain $\left( \omega_z = \frac{\partial U_y}{\partial x} - \frac{\partial U_x}{\partial y} \right)$, for the different flow-control cases in the airfoil-in-channel simulations. The detached shear layer is clearly visible in the time-averaged plot of the unactuated case. Domain clipped to exclude the sponge section.}
    \label{fig:rez_af_vortz}
\end{figure}

\begin{figure}[tb]
    \centering
    \tikzsetnextfilename{fig_af_cp_lineplot-3and6jets}
    \includegraphics{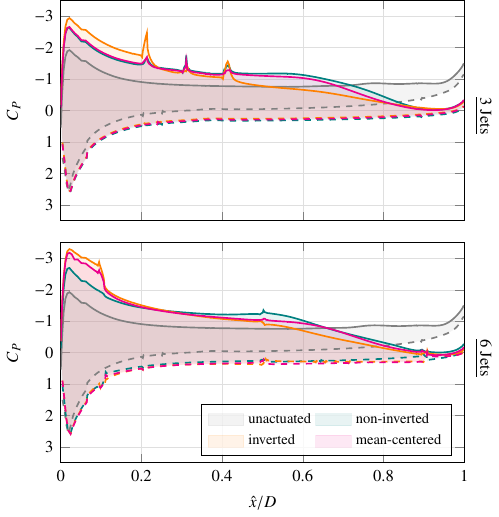}
    \caption{Time averaged pressure distributions ($C_P$ in the interval $25 \leq t^* \leq 50$) along the airfoil surfaces with and without the trained AFC agents. The dashed lines (- - -) represent the pressure surface while the solid lines (---) the suction surface. The separated flow can be observed in the flat-top $C_P$ distribution on the suction surface in the unactuated case, and the jet positions are visible as kinks in the AFC cases. The non-inverted cases (3 and 6 jets) have similar $C_P$ distributions, explaining the similarity in the response produced. Interestingly, both inverted cases \textit{weight} the $C_P$ distributions towards the leading edge.}
    \label{fig:af_cp_lineplot-3and6jets}
\end{figure}

As done previously, the best-performing policies from across the random seeds are used here for discussion.
The different AFC cases were trained on simulations lasting up to $t^*=8$ but tested on simulations lasting up to $t^*=50$, with all showing stable and consistent long-term performance.
The $C_L$, $C_D$ and $C_L/C_D$ evolution is shown in \cref{fig:rez_af_clcd,fig:rez_af_clcd-6jets} for the 3-jets and 6-jets cases respectively, while \cref{fig:rez_af_3jet_action_cost,fig:rez_af_6jet_action_cost} show the jet-strengths and running costs.
The observation from the training curves, of the non-inverted (3 and 6 jets) systems performing similarly is true in an averaged sense, but the 6-jets system has a much muted $C_D$ spread, and a much larger $C_L/C_D$ spread.
On the other hand, the 3-jets (inverted) system does look to outperform these both, with the 6-jets (inverted) system performing the best.
The mean-centered cases have higher spreads in all their respective coefficients, and in the longer-term look to perform about as well as their respective inverted counterparts.

The averaged and RMS values of the various coefficients are reported in \cref{tab:rez_af_clcdcf_latterhalf_extendtime} and they agree with the comments made so far.
The non-inverted 3-jets case increases efficiency and lift by roughly $53\%$ and $55\%$ respectively, while also increasing drag by about $2\%$.
Similarly, the non-inverted 6-jets case increases efficiency and lift by about $52\%$ and $48\%$, but reduces drag by $\sim2\%$.
The mean-centered 3-jets case performs slightly better, raising efficiency and lift by around $54\%$ and $47\%$, while cutting drag by $\sim3\%$, and has a lower averaged running cost.
The inverted 3-jets case does better than these three, raising efficiency and lift by roughly $62\%$ and $41\%$, and cutting drag by $\sim12\%$.
The 6-jets (inverted) configuration outperforms them all.
This raises efficiency by $\sim74\%$, increasing lift by roughly $49\%$ and cutting drag by $\sim14\%$, and the mean-centered case performs almost as well, but with markedly higher RMS values and cost.

As evident from the policies found in these different cases, an increase in efficiency need not rely solely on increases in lift or decreases in drag, but a suitable combination of both.
Taking a look at the RMS values, the 6-jet cases from the proposed approach both dampen drag fluctuations by over $50\%$, much more than the 3-jet cases, and also have similar lift fluctuation dampening (slightly over $30\%$).
However despite these individual fluctuation decreases, both these cases lead to an increase in fluctuation intensity in $C_L/C_D$ values, compared to the significant decreases in the 3-jet cases (roughly $-58\%$ and $-45\%$ for the non-inverted and inverted cases respectively).
Remarkably, these cases display much lower running costs as well, highlighting the effectiveness of the framework and the importance of jet-position configuration over actuation intensity.
The mean-centering cases both display much higher lift and drag fluctuation RMS values, as observed in their force evolution plots previously.

Glancing at \cref{fig:rez_af_clcd,fig:rez_af_clcd-6jets}, the lower lift and drag fluctuations in the 6-jets cases as well as the higher efficiency values in the inverted cases becomes immediately obvious.
What is further striking is the rapid pace with which the AFC systems effect their changes and increase the efficiency in all cases, and the long-term stability displayed by the policies even beyond the short training period.
Also note the quasi-periodicity that the AFC systems introduce, in a largely chaotic signal stream before their activation ($t^*<0$).

The jet-intensities and the running costs are plotted in \cref{fig:rez_af_3jet_action_cost,fig:rez_af_6jet_action_cost}.
Taking a look at the 3-jets (inverted and non-inverted) cases, the higher costs are immediately apparent, along with the strongly periodic character of the jet actuations.
In both cases, the first and third jets fire out-of-sync, whereas the second jet acts as a sort of bridge between the two.
In the non-inverted case, it eventually syncs up with the third jet, whereas in the inverted case its actuations overlap in parts with both the first and third jets.
This periodic forcing explains the periodicity observed in the $C_L$, $C_D$ and $C_L/C_D$ plots, as the AFC systems learn to create vortices rolling over the suction surface that drag in higher momentum flow from the detached shear layers closer to the airfoil.
The same quasi-periodic actuation and vortex-creation behavior is observed in the 6-jet cases, with even greater intricacy syncing up the many jets.

Surprisingly, unlike the cylinder-in-channel setup, this complex behavior is also observed in the mean-centered cases (albeit in a more chaotic fashion).
This forces us to revise our earlier comment on mean-centering hindering complex policy learning.
Given adequate care, it appears even a non-injective framework such as this can learn carefully tuned behaviors.
It does however display an affinity for higher cost expenditure when allowed, as observed in both the 4-jets cylinder and 6-jets airfoil cases, and is still non-injective in nature.

The pressure, velocity and vorticity flow-fields are shown in \cref{fig:rez_af_cp,fig:rez_af_velmag,fig:rez_af_vortz}.
The introduced periodicity is visible in the regularity of the wake in the AFC cases, compared to the chaos of the unactuated case.
The separated region is most clearly observable in \cref{fig:rez_af_velmag,fig:rez_af_vortz} in the unactuated case, along with the subsequent energization of the closest shear layer via the generated surface-hugging-vortices in the AFC cases.
\cref{fig:af_cp_lineplot-3and6jets} shows the $C_P$ distributions on the airfoil's surfaces, and makes clear once again the separated region in the unactuated case via the characteristic flat-top $C_P$ distribution on the suction surface.
The dramatic increases in lift can also be observed in these pressure plots, with the AFC systems widening the area in between the surface distributions, thereby increasing the pressure force on the airfoil.
Lastly, the jets can be observed as the pressure spikes in these plots, the \textit{spikier} 3-jet cases reflecting the higher jet actuations (and hence cost) noted previously.

\section{Conclusions}
\label{sec:conclusion}
To the best of our knowledge, this study is the first to theoretically analyze the traditional mean-centering framework for implementing zero net mass flow rate multi-jets ($N>2$, $N$ being the number of jets), having found a major overlooked mathematical flaw and derived upper bounds on suitably defined running \textit{costs}.
This maximum cost was observed to scale near-linearly with the number of jets.
Subsequently, an alternative framework remedying this flaw was proposed, from which the canonical two-jet system emerged as a special case for $N=2$.
Upper bounds on its running costs were also derived, and found to be both jet-count-independent and more economical than the mean-centering case (for $N>3$).
Two test setups, the 2D cylinder-in-channel and the 2D airfoil-in-channel environments, were used to pit the proposed multi-jet framework against the mean-centering one and see if it could actually match performance in practice.
The DRL algorithm PPO, combined with state-of-the-art practices such as KL-divergence early-stopping and learning rate warm-up, was used to train agents to reduce drag and increase aerodynamic efficiency (respectively).

Across multiple random intializations, the agents trained reliably, consistently and perhaps most importantly, quickly, thereby also reducing training compute costs.
The trained agents were able to find effective control strategies, in whichever jet-configuration they were applied to.
For the cylinder-in-channel setup with 4-jets, the agent found ways to exploit the jet positions by combining vortex shedding control with mild propulsive effects.
In the airfoil setup, the agent found ways to inject surface-hugging vortices on the suction side of the airfoil, drawing in higher momentum from the formerly separated shear layer closer to the surface and energizing the near-wall region.
In all cases across setups and configurations, the proposed approach performed at par or better than the mean-centering one, with lower costs or more muted fluctuations or both.

Thus, to conclude, the proposed framework is shown to be theoretically robust and straightforward to implement.
It removes a key mathematical flaw from the traditional case while offering similar performance in the tested setups, and possesses superior cost scaling.
Combined with best practices from standard DRL research, it consistently displays initialization independent learning, and provides a safe way to implement AFC with multiple jet actuators.

\section*{Acknowledgements}
This work was partially funded by the European Union.
It has received funding from the European High Performance Computing Joint Undertaking (JU) and Sweden, Germany, Spain, Greece, and Denmark under grant agreement No 101093393.
This research presented was funded by Deutsche For\-schungs\-ge\-mein\-schaft (DFG, German Research Foundation) under Germany's Excellence Strategy EXC 2075 -- 390740016.
We acknowledge the support by the Stuttgart Center for Simulation Science (SimTech).

\section*{Data Availability Statement}
The \relexi and \flexi codes used within this work are available under the \href{https://www.gnu.org/licenses/gpl-3.0.html}{GPLv3} license at:
\begin{itemize}
	\item \url{https://github.com/flexi-framework/relexi}
	\item \url{https://github.com/flexi-framework/flexi}
	\item \url{https://github.com/flexi-framework/flexi-extensions/tree/smartsim}
\end{itemize}

\appendix
\label{sec:app}

\section{Validation of the Simulation Setup}
\label{sec:validation}

\begin{figure}[h]
    \centering
    \tikzsetnextfilename{fig_validation-cyl}
    \includegraphics{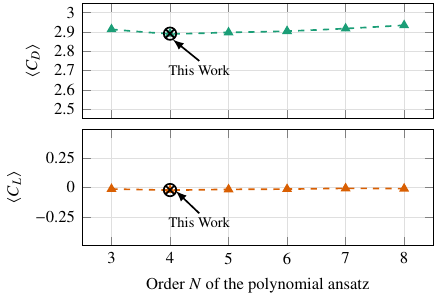}
    \caption{$C_L$ and $C_D$ variations across polynomial orders for the 2D cylinder-in-channel setup.}
    \label{fig:fig_validation-cyl}
\end{figure}

\begin{figure}[h]
    \centering
    \tikzsetnextfilename{fig_validation-af}
    \includegraphics{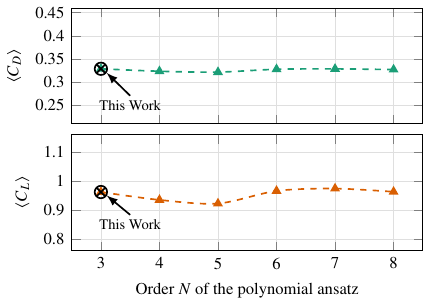}
    \caption{$C_L$ and $C_D$ variations across polynomial orders for the 2D airfoil-in-channel setup.}
    \label{fig:fig_validation-af}
\end{figure}

\cref{fig:fig_validation-cyl,fig:fig_validation-af} show the effects of polynomial order refinement on average lift and drag for the cylinder and airfoil setups respectively.
In the cylinder setup, average lift remains unchanged, whereas average drag dips and increases slightly.
In the airfoil setup, average drag remains largely unchanged, with average lift dipping and increases mildly.
Based on these plots, an order $N=4$ for the cylinder and $N=3$ for the airfoil are chosen as good trade-offs between accuracy and computational cost.
\section{Selecting Probe Locations on the Airfoil}
\label{sec:app-airfoiljetpos}

As mentioned in \cref{sec:methodology-rl}, the probe locations for the airfoil case are chosen not based on the researchers' prior experiences and biases, but rather using the deterministic heuristic described below. The only constraint applied is that the probes must be on the airfoil's surface, in order to mimic real-world limitations. The basic idea of the heuristic is to collect time-series data in the unactuated phase across a large number of probes, order them (decreasing) by their variances, and start rejecting all probes with correlations greater than a threshold $t_{corr}$. This helps remove probes with similar information content (indicated by the high correlation values), and retaining those with more distinct cues. The algorithm is given in Algorithm \ref{algo:pickprobe}:

\begin{algorithm}
    \KwData{$N_{p}$ Probes with time-series data $\mathbf{D} \in \mathbb{R}^{n_{ts} \times \ N_{p}}$, correlation threshold to reject probes $t_{corr}$}
    \KwResult{A list of probes (\texttt{selectList}) with distinct information content (correlations $<t_{corr}$)}
    \Begin{
        $\texttt{CovMat} \gets \text{CovarianceMatrix}(\mathbf{D}) \in \mathbb{R}^{N_{p}\times N_{p}}$\\
        $\texttt{CorrMat} \gets  \text{CorrelationMatrix}(\mathbf{D}) \in [-1,1]^{N_{p}\times N_{p}}$\\
        $\texttt{probeIdx} \gets \text{argsort}(\text{diag}(\texttt{CovMat})) \in \mathbb{N}^{N_p}$\\
        $\texttt{rejectList} \gets []$\\
        $\texttt{selectList} \gets []$\\
        \For{\emph{$i_p \in \rm{reversed} (\texttt{probeIdx})$}}{
            \If{\emph{$i_p \not\in \texttt{rejectList}$}}{
                $\text{Append } i_p \text{ to } \texttt{selectList}$ \\
                \For{$j_p \in \{1, \cdots, N_p\} /\{i_p\}$}{
                    \If{\emph{$\texttt{CorrMat}[i_p, j_p] \geq t_{corr}$ }}{
                        $\text{Append } j_p \text{ to } \texttt{rejectList}$
                    }
                }
            }
        }
    }
    \caption{Heuristic to pick probe locations\label{algo:pickprobe}}
\end{algorithm}

In this work, $N_p=100$ probes are initially distributed across the upper and lower surfaces of the airfoil, the pressure values are used for the time-series data $\mathbf{D}$ and $t_{corr}=0.9$ is used for the threshold. The heuristic produces a set of 28 probes, shown in \cref{fig:RPpos_airfoil}.

\section{Cost of Multi-Jet Configurations}
\label{sec:app-cost}

The cost of operating the jet-actuated AFC system is defined as in \cref{eq:cost-jetAFC}:
\begin{align}
    \mathcal{C} &= \sum_{i=1}^N |Q_i| \nonumber\\
    &= Q^{max} \sum_{i=1}^N |f_i(\mathbf{a})| . \nonumber
\end{align}
While this will vary throughout each individual simulation, a good measure to describe a given approach is to check its maximum possible cost. We do just this in the following subsections.

\subsection{Traditional Mean-Centering Approach}
\label{sec:app-cost-oldapp}

In the traditional approach to multi-jets, as described in \cref{eq:multijet-traditional} in \cref{sec:methodology-multijets}:
\begin{align}
    f_i(\mathbf{a}) = \frac{N}{(N-1)\Delta a} \left( a_i - \frac{\sum_{j=1}^N a_j}{N} \right) \nonumber
\end{align}
where $\Delta a = a_{max}-a_{min}$ and $f_i \in [-1, 1]$. Out of the $N$ jets, suppose $m$ jets are such that $f_i \geq 0$. Specifically $f_i \geq 0 \ \forall \ i \in \mathcal{P}_m$ where $|\mathcal{P}_m| = m$ and $\mathcal{P}_m \subset \mathcal{T}_N, \ \mathcal{T}_N = \{1, \cdots, N\}$ ($\mathcal{P}_m$ thus being the set of all positive $f_i$ jets). Then:
\begin{align}
    0 &= \sum_{i=1}^N f_i \nonumber \\
      &= \sum_{i\in\mathcal{P}_m} f_i + \sum_{i\in \mathcal{T}_N / \mathcal{P}_m} f_i \nonumber \\
\implies \left|  \sum_{i\in\mathcal{P}_m} f_i \right| &= \left| \sum_{i\in \mathcal{T}_N / \mathcal{P}_m} f_i \right| \label{eq:app-cost-multijet-trad-interm0}
\end{align}

From \cref{eq:meth_znmf_tradjet_bidef}:
\begin{align}
    \sum_{i\in\mathcal{P}_m} f_i &= \frac{N}{(N-1) \Delta a}\sum_{i\in\mathcal{P}_m} b_i \nonumber \\
    &= \frac{N}{(N-1) \Delta a}\sum_{i\in\mathcal{P}_m} \sum_{j=1,j\neq i}^N \left( \frac{a_i-a_j}{N} \right) \nonumber \\
    &= \frac{N}{(N-1) \Delta a}\sum_{(i,j)\in\mathcal{S}_m}  \left( \frac{a_i-a_j}{N} \right) \nonumber
\end{align}
where $\mathcal{S}_m$ is the set of index pairs defined as $\mathcal{S}_m = \{ (i,j) : i \in \mathcal{P}_m, j \in \mathcal{T}_n, j \neq i \}$.
$\mathcal{S}_m$ can then be divided into two sets $\mathcal{S}_m^1 = \{ (i,j) : i \in \mathcal{P}_m, j \in \mathcal{T}_n / \mathcal{P}_m \}$ and $\mathcal{S}_m^2 = \{ (i,j) : i \in \mathcal{P}_m, j \in \mathcal{P}_m, j \neq i \}$ such that $\mathcal{S}_m = \mathcal{S}_m^1 \cup \mathcal{S}_m^2$ and $\mathcal{S}_m^1 \cap \mathcal{S}_m^2 = \emptyset$.
Clearly,
\begin{align}
    \sum_{(i,j)\in\mathcal{S}_m^2}  \left( \frac{a_i-a_j}{N} \right) &= 0 \nonumber
\end{align}
Using this:
\begin{align}
    \sum_{i\in\mathcal{P}_m} f_i &= \frac{N}{(N-1) \Delta a}\sum_{(i,j)\in\mathcal{S}_m^1}  \left( \frac{a_i-a_j}{N} \right) \nonumber \\
    &\leq \frac{N}{(N-1) \Delta a}\sum_{(i,j)\in\mathcal{S}_m^1}  \left| \frac{a_i-a_j}{N} \right| \nonumber \\
    &\leq \frac{N}{(N-1) \Delta a} \left( \frac{m(N-m)\Delta a}{N} \right) = \frac{m(N-m)}{N-1}
\end{align}
Using \cref{eq:app-cost-multijet-trad-interm0}:
\begin{align}
    \left|  \sum_{i\in\mathcal{P}_m} f_i \right| &= \left| \sum_{i\in \mathcal{T}_N / \mathcal{P}_m} f_i \right| \leq \frac{m(N-m)}{N-1}
\end{align}

Now since $f_i \geq 0 \ \forall \ i \in \mathcal{P}_m$, $\sum_{i \in \mathcal{P}_m} |f_i| = \left| \sum_{i \in \mathcal{P}_m} f_i\right|$.
Conversely, since $f_i < 0 \ \forall \ i \in \mathcal{T}_N/\mathcal{P}_m$, $\sum_{i \in \mathcal{T}_N/\mathcal{P}_m} |f_i| = \left| \sum_{i \in \mathcal{T}_N/\mathcal{P}_m} f_i\right|$.
This leads to,
\begin{align}
    \sum_{i=1}^N |f_i| \leq \frac{2m(N-m)}{N-1} \label{eq:app-cost-multijet-trad-interm2}
\end{align}
Note that \cref{eq:app-cost-multijet-trad-interm2} holds only for `$m$' positive $f_i$. To get a generalized upper bound, we must find its maximum value across all $m$ values:
\begin{align}
    \sum_{i=1}^N |f_i| &\leq \max_m \left[\frac{2m(N-m)}{N-1}\right] \nonumber \\
    \implies \sum_{i=1}^N |f_i| &\leq  2\lfloor N/2\rfloor \left( \frac{N-\lfloor N/2\rfloor}{N-1} \right)
\end{align}
This is not merely an upper bound but an achievable upper bound, attained in the case with $m^* =\lfloor N/2 \rfloor$ jets with $a_i = a_{max}$ and $N-\lfloor N/2 \rfloor$ jets with $a_i = a_{min}$. Therefore the maximum cost of operating the AFC system in this traditional approach becomes:
\begin{align}
    C_{max} = 2 \lfloor N/2\rfloor \left( \frac{N-\lfloor N/2\rfloor}{N-1} \right) Q^{max} \label{eq:app-cost-final-oldapp}
\end{align}

\subsection{Alternative Approach}
\label{sec:app-cost-newapp}

Recall that in the new approach, as described in \cref{eq:meth-newapp-fi}, the jet contributions are determined as:
\begin{align}
    f_i(\boldsymbol{a}) &= \left( \frac{N}{N-1} \right)  b_i' \nonumber \\
    &= \begin{cases}
        \left( \frac{N}{N-1} \right) \left( \frac{a_i}{1 + \sum_{j=1}^{N-1} a_j} - \frac{1}{N} \right)  & \ \ \text{if} \ i \leq N-1, \\
        \left( \frac{N}{N-1} \right) \left( \frac{1}{1 + \sum_{j=1}^{N-1} a_j} - \frac{1}{N} \right)   & \ \ \text{if} \ i = N
    \end{cases} \nonumber
\end{align}

For computing the cost of this approach, we can use the framework detailed in the previous subsection. Noting that $f_i \in \left[\frac{-1}{N-1}, 1\right]$, we can deduce:
\begin{align}
    &0 &&\leq&& \left|  \sum_{i\in\mathcal{P}_m} f_i \right| &&\leq&& m \label{eq:app-cost-multijet-new-pbound}\\
    &0 &&<   &&    \left|  \sum_{i\in\mathcal{T}_N/\mathcal{P}_m} f_i \right| &&\leq&& (N-m) \times \frac{1}{N-1} \label{eq:app-cost-multijet-new-nbound}
\end{align}
However, since they are equal according to \cref{eq:app-cost-multijet-trad-interm0}, they must also have common bounds. This logically leads to:
\begin{align}
    \left|  \sum_{i\in\mathcal{P}_m} f_i \right| = \left|  \sum_{i\in\mathcal{T}_N/\mathcal{P}_m} f_i \right| \leq \min\left(m, \frac{N-m}{N-1} \right)
\end{align}
As shown previously, $\sum_{i \in \mathcal{P}_m} |f_i| = \left| \sum_{i \in \mathcal{P}_m} f_i\right|$ and $\sum_{i \in \mathcal{T}_N/\mathcal{P}_m} |f_i| = \left| \sum_{i \in \mathcal{T}_N/\mathcal{P}_m} f_i\right|$. Using this fact:
\begin{align}
    \sum_{i=1}^N |f_i| = 2 \left|  \sum_{i\in\mathcal{P}_m} f_i \right| \leq 2 \min\left(m, \frac{N-m}{N-1} \right) \label{eq:app-cost-multijet-new-interm2}
\end{align}

\cref{eq:app-cost-multijet-new-interm2} holds only for `$m$' positive $f_i$. To get a generalized upper bound, we must find:
\begin{align}
    \sum_{i=1}^N |f_i| = 2 \left|  \sum_{i\in\mathcal{P}_m} f_i \right| \leq 2 \max_m \left[ \min\left(m, \frac{N-m}{N-1} \right) \right]
\end{align}
Note that $\forall \ m > 1$, $\frac{N-m}{N-1} < 1$ and therefore $\min\left(m, \frac{N-m}{N-1}\right) = \frac{N-m}{N-1} < 1$. For $m = 1$, $\min\left(m, \frac{N-m}{N-1}\right) = 1$. Therefore:
\begin{align}
    \sum_{i=1}^N |f_i| \leq 2
\end{align}
This too is not merely an upper bound but an achievable upper bound, attained in the case with $m^*=1$ jet set to $f_i = 1$ and the remaining $N-1$ jets set to $f_i = \frac{-1}{N-1}$. Therefore the maximum cost of operating the AFC system in this alternative approach becomes:
\begin{align}
    C_{max} = 2 Q^{max} \label{eq:app-cost-final-newapp}
\end{align}

\bibliographystyle{unsrtnat}

\begin{thebibliography}{61}
\providecommand{\natexlab}[1]{#1}
\providecommand{\url}[1]{\texttt{#1}}
\expandafter\ifx\csname urlstyle\endcsname\relax
  \providecommand{\doi}[1]{doi: #1}\else
  \providecommand{\doi}{doi: \begingroup \urlstyle{rm}\Url}\fi

\bibitem[Gad-el Hak(2000)]{Gad-el-Hak_2000}
Mohamed Gad-el Hak.
\newblock \emph{Flow Control: Passive, Active, and Reactive Flow Management}.
\newblock Cambridge University Press, 2000.

\bibitem[McLellan and Ladson(1988)]{mclellan1988history}
Bruce~W. McLellan and Charles~L. Ladson.
\newblock A history of suction-type laminar-flow control with emphasis on
  flight research.
\newblock Technical Report TM-4080, NASA, 1988.

\bibitem[Beratlis et~al.(2017)Beratlis, Squires, and
  Balaras]{beratlis_separation_2017}
Nikolaos Beratlis, Kyle~D. Squires, and Elias Balaras.
\newblock Separation control and drag reduction using roughness elements.
\newblock In \emph{Proceeding of {Tenth} {International} {Symposium} on
  {Turbulence} and {Shear} {Flow} {Phenomena}}, pages 199--204. Begellhouse,
  2017.
\newblock Place: Swissotel Chicago, Chicago, Illinois, U.S.A.

\bibitem[Bechert and Bartenwerfer(1989)]{bechert_viscous_1989}
D.~W. Bechert and M.~Bartenwerfer.
\newblock The viscous flow on surfaces with longitudinal ribs.
\newblock \emph{Journal of Fluid Mechanics}, 206:\penalty0 105--129, September
  1989.
\newblock ISSN 1469-7645, 0022-1120.

\bibitem[Chambers(2003)]{Chambers2003ConceptTR}
Joseph~R Chambers.
\newblock Concept to reality: contributions of the langley research center to
  us civil aircraft of the 1990s.
\newblock \emph{NASA Special Publication}, page 59513, 2003.

\bibitem[Lin(2002)]{lin2002review}
John~C Lin.
\newblock Review of research on low-profile vortex generators to control
  boundary-layer separation.
\newblock \emph{Progress in aerospace sciences}, 38\penalty0 (4-5):\penalty0
  389--420, 2002.

\bibitem[Amitay et~al.(1998)Amitay, Smith, and Glezer]{amitay1998aerodynamic}
Michael Amitay, Barton Smith, and Ari Glezer.
\newblock Aerodynamic flow control using synthetic jet technology.
\newblock In \emph{36th AIAA Aerospace Sciences Meeting and Exhibit}, page 208,
  1998.

\bibitem[Greenblatt and Wygnanski(2000)]{greenblatt_control_2000}
David Greenblatt and Israel~J. Wygnanski.
\newblock The control of flow separation by periodic excitation.
\newblock \emph{Progress in Aerospace Sciences}, 36\penalty0 (7):\penalty0
  487--545, October 2000.
\newblock ISSN 0376-0421.

\bibitem[Kametani and Fukagata(2011)]{kametani_direct_2011}
Yukinori Kametani and Koji Fukagata.
\newblock Direct numerical simulation of spatially developing turbulent
  boundary layers with uniform blowing or suction.
\newblock \emph{Journal of Fluid Mechanics}, 681:\penalty0 154--172, August
  2011.
\newblock ISSN 1469-7645, 0022-1120.

\bibitem[Yousefi and Saleh(2015)]{yousefi_three-dimensional_2015}
Kianoosh Yousefi and Reza Saleh.
\newblock Three-dimensional suction flow control and suction jet length
  optimization of {NACA} 0012 wing.
\newblock \emph{Meccanica}, 50\penalty0 (6):\penalty0 1481--1494, June 2015.
\newblock ISSN 1572-9648.

\bibitem[Voevodin et~al.(2019)Voevodin, Kornyakov, Petrov, Petrov, and
  Sudakov]{voevodin_improvement_2019}
A.~V. Voevodin, A.~A. Kornyakov, A.~S. Petrov, D.~A. Petrov, and G.~G. Sudakov.
\newblock Improvement of the take-off and landing characteristics of wing using
  an ejector pump.
\newblock \emph{Thermophysics and Aeromechanics}, 26\penalty0 (1):\penalty0
  9--18, January 2019.
\newblock ISSN 1531-8699.

\bibitem[Desalvo et~al.(2012)Desalvo, Whalen, and Glezer]{desalvo2012high}
Michael Desalvo, Edward Whalen, and Ari Glezer.
\newblock High-lift enhancement using active flow control.
\newblock In \emph{6th AIAA flow control conference}, page 3245, 2012.

\bibitem[Wu et~al.(1998)Wu, Lu, Denny, Fan, and Wu]{wu1998post}
Jie-Zhi Wu, Xi-Yun Lu, Andrew~G Denny, Meng Fan, and Jain-Ming Wu.
\newblock Post-stall flow control on an airfoil by local unsteady forcing.
\newblock \emph{Journal of fluid Mechanics}, 371:\penalty0 21--58, 1998.

\bibitem[Radespiel et~al.(2016)Radespiel, Burnazzi, Casper, and
  Scholz]{radespiel2016active}
Rolf Radespiel, Marco Burnazzi, M~Casper, and P~Scholz.
\newblock Active flow control for high lift with steady blowing.
\newblock \emph{The Aeronautical Journal}, 120\penalty0 (1223):\penalty0
  171--200, 2016.

\bibitem[Warui and Fujisawa(1996)]{warui1996feedback}
HM~Warui and N~Fujisawa.
\newblock Feedback control of vortex shedding from a circular cylinder by
  cross-flow cylinder oscillations.
\newblock \emph{Experiments in Fluids}, 21\penalty0 (1):\penalty0 49--56, 1996.

\bibitem[Min and Choi(1999)]{min1999suboptimal}
Chulhong Min and Haecheon Choi.
\newblock Suboptimal feedback control of vortex shedding at low reynolds
  numbers.
\newblock \emph{Journal of Fluid Mechanics}, 401:\penalty0 123--156, 1999.

\bibitem[Muddada and Patnaik(2010)]{muddada_active_2010}
Sridhar Muddada and B.~S.~V. Patnaik.
\newblock An active flow control strategy for the suppression of vortex
  structures behind a circular cylinder.
\newblock \emph{European Journal of Mechanics - B/Fluids}, 29\penalty0
  (2):\penalty0 93--104, March 2010.
\newblock ISSN 0997-7546.

\bibitem[Lee et~al.(2013)Lee, Kim, Choi, Kim, Kim, and Jung]{lee2013closed}
Byunghyun Lee, Minhee Kim, Byounghun Choi, Chongam Kim, H~Jin Kim, and
  Kyoung~Jin Jung.
\newblock Closed-loop active flow control of stall separation using synthetic
  jets.
\newblock In \emph{31st AIAA applied aerodynamics conference}, page 2925, 2013.

\bibitem[Michel et~al.(2024)Michel, Neunaber, Mishra, Braud, Plestan, Barbot,
  and Hamon]{michel2024novel}
Lo{\"\i}c Michel, Ingrid Neunaber, Rishabh Mishra, Caroline Braud, Franck
  Plestan, Jean-Pierre Barbot, and Pol Hamon.
\newblock A novel lift controller for a wind turbine blade section using an
  active flow control device including saturations: experimental results.
\newblock \emph{IEEE Transactions on Control Systems Technology}, 32\penalty0
  (5):\penalty0 1590--1601, 2024.

\bibitem[Kaul(2022)]{kaul2022active}
Upender~K Kaul.
\newblock An active flow control approach for spatially growing mixing layer.
\newblock \emph{Journal of Fluids Engineering}, 144\penalty0 (6):\penalty0
  061110, 2022.

\bibitem[Kiesner and King(2017)]{doi:10.2514/1.J055728}
Matthias Kiesner and Rudibert King.
\newblock Multivariable closed-loop active flow control of a compressor stator
  cascade.
\newblock \emph{AIAA Journal}, 55\penalty0 (10):\penalty0 3371--3380, 2017.
\newblock \doi{10.2514/1.J055728}.
\newblock URL \url{https://doi.org/10.2514/1.J055728}.

\bibitem[Staats et~al.(2017)Staats, Nitsche, Steinberg, and
  King]{staats2017closed}
Marcel Staats, W~Nitsche, SJ~Steinberg, and R~King.
\newblock Closed-loop active flow control of a non-steady flow field in a
  highly-loaded compressor cascade.
\newblock \emph{CEAS Aeronautical Journal}, 8\penalty0 (1):\penalty0 197--208,
  2017.

\bibitem[Krentel et~al.(2010)Krentel, Muminovic, Brunn, Nitsche, and
  King]{krentel2010application}
Daniel Krentel, Rifet Muminovic, Andr{\'e} Brunn, Wolfgang Nitsche, and
  Rudibert King.
\newblock Application of active flow control on generic 3d car models.
\newblock In \emph{Active Flow Control II: Papers Contributed to the
  Conference” Active Flow Control II 2010”, Berlin, Germany, May 26 to 28,
  2010}, pages 223--239. Springer, 2010.

\bibitem[Rabault et~al.(2019)Rabault, Kuchta, Jensen, Reglade, and
  Cerardi]{rabault2019artificial}
Jean Rabault, Miroslav Kuchta, Atle Jensen, Ulysse Reglade, and Nicolas
  Cerardi.
\newblock Artificial neural networks trained through deep reinforcement
  learning discover control strategies for active flow control.
\newblock \emph{Journal of Fluid Mechanics}, 865:\penalty0 281--302, apr 2019.
\newblock ISSN 0022-1120, 1469-7645.
\newblock \doi{10.1017/jfm.2019.62}.

\bibitem[Rabault and Kuhnle(2019)]{rabault2019accelerating}
Jean Rabault and Alexander Kuhnle.
\newblock Accelerating deep reinforcement learning strategies of flow control
  through a multi-environment approach.
\newblock \emph{Physics of Fluids}, 31\penalty0 (9):\penalty0 094105, sep 2019.
\newblock ISSN 1070-6631, 1089-7666.
\newblock \doi{10.1063/1.5116415}.

\bibitem[Rabault et~al.(2020)Rabault, Ren, Zhang, Tang, and
  Xu]{rabault2020deep}
Jean Rabault, Feng Ren, Wei Zhang, Hui Tang, and Hui Xu.
\newblock Deep reinforcement learning in fluid mechanics: a promising method
  for both active flow control and shape optimization.
\newblock \emph{arXiv preprint arXiv:2001.02464}, 2020.

\bibitem[Tang et~al.(2020)Tang, Rabault, Kuhnle, Wang, and
  Wang]{tang2020robust}
Hongwei Tang, Jean Rabault, Alexander Kuhnle, Yan Wang, and Tongguang Wang.
\newblock Robust active flow control over a range of reynolds numbers using an
  artificial neural network trained through deep reinforcement learning.
\newblock \emph{Physics of Fluids}, 32\penalty0 (5):\penalty0 053605, 05 2020.
\newblock ISSN 1070-6631.
\newblock \doi{10.1063/5.0006492}.
\newblock URL \url{https://doi.org/10.1063/5.0006492}.

\bibitem[Wang et~al.(2022)Wang, Mei, Aubry, Chen, Wu, and Wu]{wang2022deep}
Yi-Zhe Wang, Yu-Fei Mei, Nadine Aubry, Zhihua Chen, Peng Wu, and Wei-Tao Wu.
\newblock Deep reinforcement learning based synthetic jet control on disturbed
  flow over airfoil.
\newblock \emph{Physics of Fluids}, 34\penalty0 (3), 2022.

\bibitem[Garcia et~al.(2025)Garcia, Miró, Suárez, Alcántara-Ávila, Rabault,
  Font, Lehmkuhl, and Vinuesa]{GARCIA2025109913}
Xavier Garcia, Arnau Miró, Pol Suárez, Francisco Alcántara-Ávila, Jean
  Rabault, Bernat Font, Oriol Lehmkuhl, and Ricardo Vinuesa.
\newblock Deep-reinforcement-learning-based separation control in a
  two-dimensional airfoil.
\newblock \emph{International Journal of Heat and Fluid Flow}, 116:\penalty0
  109913, 2025.
\newblock ISSN 0142-727X.
\newblock \doi{https://doi.org/10.1016/j.ijheatfluidflow.2025.109913}.
\newblock URL
  \url{https://www.sciencedirect.com/science/article/pii/S0142727X25001717}.

\bibitem[Su{\'a}rez et~al.(2023)Su{\'a}rez, Alc{\'a}ntara-{\'A}vila, Mir{\'o},
  Rabault, Font, and Lehmkuhl]{suarez2023active}
Pol Su{\'a}rez, Francisco Alc{\'a}ntara-{\'A}vila, Arnau Mir{\'o}, Jean
  Rabault, Bernat Font, and Oriol Lehmkuhl.
\newblock Active flow control for three-dimensional cylinders through deep
  reinforcement learning.
\newblock \emph{arXiv preprint arXiv:2309.02462}, 2023.

\bibitem[Su{\'a}rez et~al.(2024)Su{\'a}rez, {\'A}lcantara-{\'A}vila, Rabault,
  Mir{\'o}, Font, Lehmkuhl, and Vinuesa]{suarez2024flow}
P.~Su{\'a}rez, F.~{\'A}lcantara-{\'A}vila, J.~Rabault, A.~Mir{\'o}, B.~Font,
  O.~Lehmkuhl, and R.~Vinuesa.
\newblock Flow control of three-dimensional cylinders transitioning to
  turbulence via multi-agent reinforcement learning, may 2024.

\bibitem[Su{\'a}rez et~al.(2025{\natexlab{a}})Su{\'a}rez,
  Alc{\'a}ntara-{\'A}vila, Mir{\'o}, Rabault, Font, Lehmkuhl, and
  Vinuesa]{suarez2025active}
Pol Su{\'a}rez, Francisco Alc{\'a}ntara-{\'A}vila, Arnau Mir{\'o}, Jean
  Rabault, Bernat Font, Oriol Lehmkuhl, and Ricardo Vinuesa.
\newblock Active flow control for drag reduction through multi-agent
  reinforcement learning on a turbulent cylinder at r e d= 3900.
\newblock \emph{Flow, Turbulence and Combustion}, pages 1--25,
  2025{\natexlab{a}}.

\bibitem[Mondal et~al.(2025)Mondal, Vinuesa, and Jagtap]{mondal2025shocks}
Trishit Mondal, Ricardo Vinuesa, and Ameya~D Jagtap.
\newblock Shocks under control: Taming transonic compressible flow over an
  rae2822 airfoil with deep reinforcement learning.
\newblock \emph{arXiv preprint arXiv:2511.07564}, 2025.

\bibitem[Su{\'a}rez et~al.(2025{\natexlab{b}})Su{\'a}rez,
  {Alc{\'a}ntara-{\'A}vila}, Mir{\'o}, Rabault, Font, Lehmkuhl, and
  Vinuesa]{suarez2025flow}
Pol Su{\'a}rez, Francisco {Alc{\'a}ntara-{\'A}vila}, Arnau Mir{\'o}, Jean
  Rabault, Bernat Font, Oriol Lehmkuhl, and Ricardo Vinuesa.
\newblock Active flow control for drag reduction through multi-agent
  reinforcement learning on a turbulent cylinder at $\mathrm{Re}_{D}=3900$.
\newblock \emph{Flow, Turbulence and Combustion}, mar 2025{\natexlab{b}}.
\newblock ISSN 1386-6184, 1573-1987.
\newblock \doi{10.1007/s10494-025-00642-x}.

\bibitem[Kurz et~al.(2025)Kurz, Kaushik, Blind, Kopper, Schwarz, Rodach, and
  Beck]{KURZ2025106854}
Marius Kurz, Rohan Kaushik, Marcel Blind, Patrick Kopper, Anna Schwarz, Felix
  Rodach, and Andrea Beck.
\newblock Invariant control strategies for active flow control using graph
  neural networks.
\newblock \emph{Computers \& Fluids}, 303:\penalty0 106854, 2025.
\newblock ISSN 0045-7930.
\newblock \doi{https://doi.org/10.1016/j.compfluid.2025.106854}.
\newblock URL
  \url{https://www.sciencedirect.com/science/article/pii/S0045793025003147}.

\bibitem[Carlson(2011)]{carlson2011FUN3D}
Jan-Rene{\'e} Carlson.
\newblock Inflow/outflow boundary conditions with application to {{FUN3D}}.
\newblock Technical Report NASA/TM--2011-217181, Langley Research Center,
  Langley Research Center, Hampton, VA, United States, oct 2011.

\bibitem[Krais et~al.(2021)Krais, Beck, Bolemann, Frank, Flad, Gassner,
  Hindenlang, Hoffmann, Kuhn, Sonntag, and Munz]{krais2021flexi}
Nico Krais, Andrea Beck, Thomas Bolemann, Hannes Frank, David Flad, Gregor
  Gassner, Florian Hindenlang, Malte Hoffmann, Thomas Kuhn, Matthias Sonntag,
  and Claus-Dieter Munz.
\newblock {{FLEXI}}: {{A}} high order discontinuous {{Galerkin}} framework for
  hyperbolic--parabolic conservation laws.
\newblock \emph{Computers \& Mathematics with Applications}, 81:\penalty0
  186--219, jan 2021.
\newblock ISSN 08981221.
\newblock \doi{10.1016/j.camwa.2020.05.004}.

\bibitem[Blind et~al.(2024{\natexlab{a}})Blind, Kopper, Kempf, Kurz, Schwarz,
  Munz, and Beck]{blind2022performance}
Marcel Blind, Patrick Kopper, Daniel Kempf, Marius Kurz, Anna Schwarz,
  Claus-Dieter Munz, and Andrea Beck.
\newblock \emph{Performance Improvements for Large Scale Simulations Using the
  Discontinuous Galerkin Framework FLEXI}, pages 249--264.
\newblock Springer Nature Switzerland, Cham, 2024{\natexlab{a}}.
\newblock ISBN 978-3-031-46870-4.

\bibitem[Blind et~al.(2024{\natexlab{b}})Blind, Gibis, Wenzel, and
  Beck]{blind2024wall}
Marcel~P. Blind, Tobias Gibis, Christoph Wenzel, and Andrea Beck.
\newblock {Wall-modeled large eddy simulation of a tandem wing configuration in
  transonic flow}.
\newblock \emph{Physics of Fluids}, 36\penalty0 (5):\penalty0 055125, 05
  2024{\natexlab{b}}.
\newblock ISSN 1070-6631.
\newblock \doi{10.1063/5.0198271}.

\bibitem[Dürrwächter et~al.(2021)Dürrwächter, Kurz, Kopper, Kempf, Munz,
  and Beck]{durrwachter2021efficient}
Jakob Dürrwächter, Marius Kurz, Patrick Kopper, Daniel Kempf, Claus-Dieter
  Munz, and Andrea Beck.
\newblock An efficient sliding mesh interface method for high-order
  discontinuous galerkin schemes.
\newblock \emph{Computers \& Fluids}, 217:\penalty0 104825, 2021.
\newblock ISSN 0045-7930.
\newblock \doi{https://doi.org/10.1016/j.compfluid.2020.104825}.
\newblock URL
  \url{https://www.sciencedirect.com/science/article/pii/S0045793020303959}.

\bibitem[Schulman et~al.(2017)Schulman, Wolski, Dhariwal, Radford, and
  Klimov]{schulman2017proximal}
John Schulman, Filip Wolski, Prafulla Dhariwal, Alec Radford, and Oleg Klimov.
\newblock Proximal policy optimization algorithms.
\newblock \emph{arXiv preprint}, 2017.

\bibitem[Lillicrap et~al.(2015)Lillicrap, Hunt, Pritzel, Heess, Erez, Tassa,
  Silver, and Wierstra]{lillicrap2015continuous}
Timothy~P Lillicrap, Jonathan~J Hunt, Alexander Pritzel, Nicolas Heess, Tom
  Erez, Yuval Tassa, David Silver, and Daan Wierstra.
\newblock Continuous control with deep reinforcement learning.
\newblock \emph{arXiv preprint arXiv:1509.02971}, 2015.

\bibitem[Barth-Maron et~al.(2018)Barth-Maron, Hoffman, Budden, Dabney, Horgan,
  Tb, Muldal, Heess, and Lillicrap]{barth2018distributed}
Gabriel Barth-Maron, Matthew~W Hoffman, David Budden, Will Dabney, Dan Horgan,
  Dhruva Tb, Alistair Muldal, Nicolas Heess, and Timothy Lillicrap.
\newblock Distributed distributional deterministic policy gradients.
\newblock \emph{arXiv preprint arXiv:1804.08617}, 2018.

\bibitem[Gu et~al.(2016)Gu, Lillicrap, Sutskever, and Levine]{gu2016continuous}
Shixiang Gu, Timothy Lillicrap, Ilya Sutskever, and Sergey Levine.
\newblock Continuous deep {Q}-learning with model-based acceleration.
\newblock In Maria~Florina Balcan and Kilian~Q. Weinberger, editors,
  \emph{Proceedings of The 33rd International Conference on Machine Learning},
  volume~48 of \emph{Proceedings of Machine Learning Research}, pages
  2829--2838, New York, New York, USA, 20--22 Jun 2016. PMLR.
\newblock URL \url{https://proceedings.mlr.press/v48/gu16.html}.

\bibitem[Sutton et~al.(2020)Sutton, Barto, et~al.]{sutton2020reinforcement}
Richard~S Sutton, Andrew~G Barto, et~al.
\newblock \emph{Reinforcement learning: An introduction}, volume~1.
\newblock MIT press Cambridge, 2020.

\bibitem[Lax(1973)]{lax1973hyperbolic}
Peter~D Lax.
\newblock \emph{Hyperbolic systems of conservation laws and the mathematical
  theory of shock waves}.
\newblock SIAM, 1973.

\bibitem[Toro(2013)]{toro2013riemann}
Eleuterio~F Toro.
\newblock \emph{Riemann solvers and numerical methods for fluid dynamics: a
  practical introduction}.
\newblock Springer Science \& Business Media, 2013.

\bibitem[Kurz et~al.(2022)Kurz, Offenh{\"a}user, Viola, Resch, and
  Beck]{kurz2022relexi}
Marius Kurz, Philipp Offenh{\"a}user, Dominic Viola, Michael Resch, and Andrea
  Beck.
\newblock Relexi --- {{A}} scalable open source reinforcement learning
  framework for high-performance computing.
\newblock \emph{Software Impacts}, 14:\penalty0 100422, dec 2022.
\newblock ISSN 2665-9638.
\newblock \doi{10.1016/j.simpa.2022.100422}.

\bibitem[Abadi et~al.(2015)Abadi, Agarwal, Barham, Brevdo, Chen, Citro,
  Corrado, Davis, Dean, Devin, Ghemawat, Goodfellow, Harp, Irving, Isard, Jia,
  Jozefowicz, Kaiser, Kudlur, Levenberg, Man\'{e}, Monga, Moore, Murray, Olah,
  Schuster, Shlens, Steiner, Sutskever, Talwar, Tucker, Vanhoucke, Vasudevan,
  Vi\'{e}gas, Vinyals, Warden, Wattenberg, Wicke, Yu, and
  Zheng]{tensorflow2015-whitepaper}
Mart\'{i}n Abadi, Ashish Agarwal, Paul Barham, Eugene Brevdo, Zhifeng Chen,
  Craig Citro, Greg~S. Corrado, Andy Davis, Jeffrey Dean, Matthieu Devin,
  Sanjay Ghemawat, Ian Goodfellow, Andrew Harp, Geoffrey Irving, Michael Isard,
  Yangqing Jia, Rafal Jozefowicz, Lukasz Kaiser, Manjunath Kudlur, Josh
  Levenberg, Dandelion Man\'{e}, Rajat Monga, Sherry Moore, Derek Murray, Chris
  Olah, Mike Schuster, Jonathon Shlens, Benoit Steiner, Ilya Sutskever, Kunal
  Talwar, Paul Tucker, Vincent Vanhoucke, Vijay Vasudevan, Fernanda Vi\'{e}gas,
  Oriol Vinyals, Pete Warden, Martin Wattenberg, Martin Wicke, Yuan Yu, and
  Xiaoqiang Zheng.
\newblock {TensorFlow}: Large-scale machine learning on heterogeneous systems,
  2015.
\newblock URL \url{https://www.tensorflow.org/}.
\newblock Software available from tensorflow.org.

\bibitem[Guadarrama et~al.(2018)Guadarrama, Korattikara, Ramirez, Castro,
  Holly, Fishman, Wang, Gonina, Wu, Kokiopoulou, Sbaiz, Smith, Bartók, Berent,
  Harris, Vanhoucke, and Brevdo]{TFAgents}
Sergio Guadarrama, Anoop Korattikara, Oscar Ramirez, Pablo Castro, Ethan Holly,
  Sam Fishman, Ke~Wang, Ekaterina Gonina, Neal Wu, Efi Kokiopoulou, Luciano
  Sbaiz, Jamie Smith, Gábor Bartók, Jesse Berent, Chris Harris, Vincent
  Vanhoucke, and Eugene Brevdo.
\newblock {TF-Agents}: A library for reinforcement learning in tensorflow.
\newblock \url{https://github.com/tensorflow/agents}, 2018.
\newblock URL \url{https://github.com/tensorflow/agents}.
\newblock [Online; accessed 25-June-2019].

\bibitem[Partee et~al.(2022)Partee, Ellis, Rigazzi, Shao, Bachman, Marques, and
  Robbins]{partee2022using}
Sam Partee, Matthew Ellis, Alessandro Rigazzi, Andrew~E. Shao, Scott Bachman,
  Gustavo Marques, and Benjamin Robbins.
\newblock Using machine learning at scale in numerical simulations with
  {{SmartSim}}: {{An}} application to ocean climate modeling.
\newblock \emph{Journal of Computational Science}, 62:\penalty0 101707, July
  2022.
\newblock ISSN 18777503.
\newblock \doi{10.1016/j.jocs.2022.101707}.

\bibitem[Engstrom et~al.(2019)Engstrom, Ilyas, Santurkar, Tsipras, Janoos,
  Rudolph, and Madry]{engstrom2019implementation}
Logan Engstrom, Andrew Ilyas, Shibani Santurkar, Dimitris Tsipras, Firdaus
  Janoos, Larry Rudolph, and Aleksander Madry.
\newblock Implementation matters in deep {RL}: A case study on {PPO} and
  {TRPO}.
\newblock In \emph{International conference on learning representations}, 2019.

\bibitem[Kingma and Ba(2014)]{kingma2014adam}
Diederik~P Kingma and Jimmy Ba.
\newblock Adam: A method for stochastic optimization.
\newblock \emph{arXiv preprint arXiv:1412.6980}, 2014.

\bibitem[Liu et~al.(2025)Liu, Ge, Pan, Kang, and Zhang]{liu2025theoretical}
Yuxing Liu, Yuze Ge, Rui Pan, An~Kang, and Tong Zhang.
\newblock Theoretical analysis on how learning rate warmup accelerates
  convergence.
\newblock \emph{arXiv preprint arXiv:2509.07972}, 2025.

\bibitem[Abuduweili and Liu(2024)]{abuduweili2024revisiting}
Abulikemu Abuduweili and Changliu Liu.
\newblock Revisiting the initial steps in adaptive gradient descent
  optimization.
\newblock \emph{arXiv preprint arXiv:2412.02153}, 2024.

\bibitem[Kalra and Barkeshli(2024)]{kalra2024warmup}
Dayal~Singh Kalra and Maissam Barkeshli.
\newblock Why warmup the learning rate? underlying mechanisms and improvements.
\newblock \emph{Advances in Neural Information Processing Systems},
  37:\penalty0 111760--111801, 2024.

\bibitem[Vaswani et~al.(2017)Vaswani, Shazeer, Parmar, Uszkoreit, Jones, Gomez,
  Kaiser, and Polosukhin]{vaswani2017attention}
Ashish Vaswani, Noam Shazeer, Niki Parmar, Jakob Uszkoreit, Llion Jones,
  Aidan~N Gomez, {\L}ukasz Kaiser, and Illia Polosukhin.
\newblock Attention is all you need.
\newblock \emph{Advances in neural information processing systems}, 30, 2017.

\bibitem[Goyal et~al.(2017)Goyal, Doll{\'a}r, Girshick, Noordhuis, Wesolowski,
  Kyrola, Tulloch, Jia, and He]{goyal2017accurate}
Priya Goyal, Piotr Doll{\'a}r, Ross Girshick, Pieter Noordhuis, Lukasz
  Wesolowski, Aapo Kyrola, Andrew Tulloch, Yangqing Jia, and Kaiming He.
\newblock Accurate, large minibatch sgd: Training imagenet in 1 hour.
\newblock \emph{arXiv preprint arXiv:1706.02677}, 2017.

\bibitem[Gotmare et~al.(2018)Gotmare, Keskar, Xiong, and
  Socher]{gotmare2018closer}
Akhilesh Gotmare, Nitish~Shirish Keskar, Caiming Xiong, and Richard Socher.
\newblock A closer look at deep learning heuristics: Learning rate restarts,
  warmup and distillation.
\newblock \emph{arXiv preprint arXiv:1810.13243}, 2018.

\bibitem[Vignon et~al.(2023)Vignon, Rabault, and Vinuesa]{vignon2023recent}
C.~Vignon, J.~Rabault, and R.~Vinuesa.
\newblock Recent advances in applying deep reinforcement learning for flow
  control: {{Perspectives}} and future directions.
\newblock \emph{Physics of Fluids}, 35\penalty0 (3):\penalty0 031301, mar 2023.
\newblock ISSN 1070-6631, 1089-7666.
\newblock \doi{10.1063/5.0143913}.

\bibitem[Vinuesa(2024)]{vinuesa2024perspectives}
Ricardo Vinuesa.
\newblock Perspectives on predicting and controlling turbulent flows through
  deep learning.
\newblock \emph{Physics of Fluids}, 36\penalty0 (3):\penalty0 031401, March
  2024.
\newblock ISSN 1070-6631, 1089-7666.
\newblock \doi{10.1063/5.0190452}.

\end{thebibliography}

\end{document}